\documentclass[]{fairmeta}
\usepackage{times}

\usepackage{amsmath,amsfonts,bm}

\def\eqref#1{equation~\ref{#1}}
\def\1{\bm{1}}

\DeclareMathAlphabet{\mathsfit}{\encodingdefault}{\sfdefault}{m}{sl}
\SetMathAlphabet{\mathsfit}{bold}{\encodingdefault}{\sfdefault}{bx}{n}

\usepackage{hyperref}
\usepackage{url}

\usepackage{multirow}
\usepackage{tabularx}
\usepackage{array}
\usepackage{colortbl}
\usepackage{xcolor}
\usepackage{soul}
\usepackage{makecell}
\usepackage{seqsplit}
\usepackage{graphicx}
\usepackage{wrapfig}
\usepackage{booktabs}
\usepackage{pifont}
\usepackage{amssymb}
\usepackage{amsmath}
\usepackage{caption}
\usepackage{graphicx}
\usepackage{pifont}
\usepackage{enumitem}
\usepackage{float}
\usepackage{algorithm, algpseudocode}
\newcommand{\yes}{\ding{51}} % ✓
\newcommand{\no}{\ding{55}}

\usepackage{subcaption}
\usepackage{xspace}
\usepackage{wrapfig,lipsum,booktabs}
\definecolor{LightCyan}{rgb}{0.88,1,1}
\definecolor{lgreen}{HTML}{D5E8D4}
\definecolor{green}{HTML}{77B254}

\hypersetup{hidelinks}

\newcommand{\tte}{\ensuremath{\mathrm{TTE}}\xspace}
\newcommand{\ours}{\ensuremath{\mathrm{TTE}\text{-v2}}\xspace}
\newcommand{\ecrr}{\ensuremath{\mathrm{ECRR}}\xspace}
\newcommand{\qar}{\ensuremath{\mathrm{QAR}}\xspace}
\newcommand{\hnm}{\ensuremath{\mathrm{rHNM}}\xspace}
\newcommand{\pls}[1]{\textcolor{green}{\scriptsize\,(+#1)}}
\newcommand{\mns}[1]{\textcolor{red}{\scriptsize\,(-#1)}}

\title{Reason to Contrast: A Cascaded Multimodal Retrieval Framework}

\author[2, *, \dagger]{Xuanming Cui}
\author[1, \dagger]{Hong-you Chen}
\author[1]{Hao Yu}
\author[1]{Hao Yuan}
\author[1]{Zihao Wang}
\author[1]{Shlok Kumar Mishra}
\author[1]{Hanchao Yu}
\author[1]{Yonghuan Yang}
\author[1]{Jun Xiao}
\author[2]{Ser-Nam Lim}
\author[1]{Jianpeng Cheng}
\author[1]{Qi Guo}
\author[1]{Xiangjun Fan}

\affiliation[1]{AI at Meta}
\affiliation[2]{University of Central Florida}

\contribution[*]{Work done at Meta}
\contribution[\dagger]{Joint first author}
\abstract{\begin{abstract}
Traditional multimodal retrieval systems rely primarily on bi-encoder architectures, where performance is closely tied to embedding dimensionality. A recent work Think-Then-Embed (TTE) shows that incorporating multimodal reasoning to elicit additional informative tokens before embedding can further improve retrieval. In this paper, we extend this paradigm with \ours, a hybrid multimodal retrieval framework that introduces reasoning-driven performance scaling based on additional input token budget rather than model or embedding size. Our approach augments the initial multimodal retrieval with additional reasoning steps for reranking, enabling more expressive query–candidate interactions at test time. 
The reranking stage further provides finegrained supervision for hard negative mining and false negative filtering, creating a feedback loop that effectively strengthens the upstream retriever. This cascaded design delivers substantial test-time improvements based on intermediate reasoning token scaling. Experiments on MMEB-V2 benchmark demonstrate that \ours-7B achieves a new state of the art with 75.7\% accuracy, and that \ours-2B matches or surpasses leading 7B models trained with significantly larger external data. Our results highlight the promise of token-wise scaling as an alternative scaling view for multimodal retrieval.

\end{abstract}}
\date{\today}
\begin{document}

%\mymaketitle
\maketitle

\section{Introduction}
\label{sec:intro}

\begin{wrapfigure}{r}{0.5\textwidth}
    \vspace{-15pt}
    \centering
    \includegraphics[width=\linewidth]{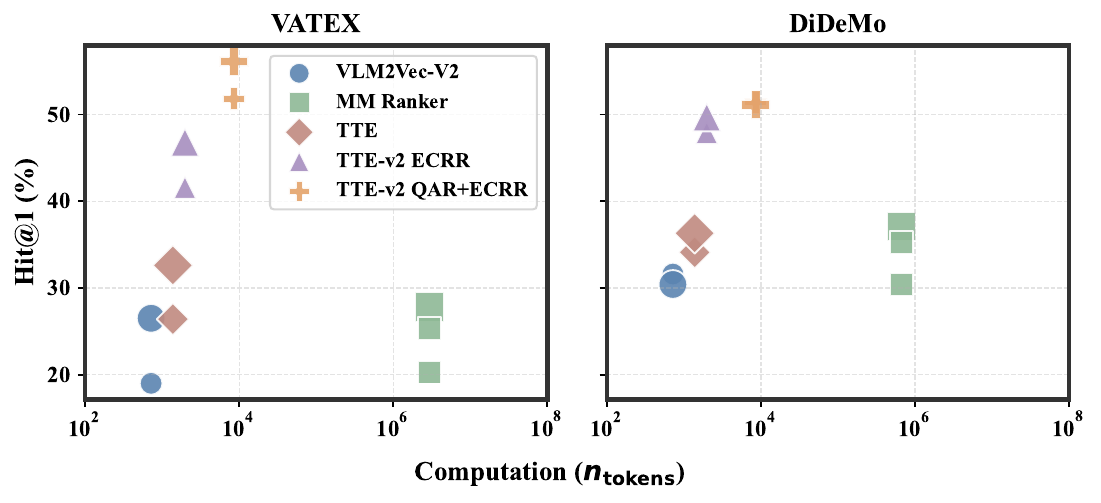}
    \vspace{-15pt}
    \caption{\footnotesize \textbf{TTE-v2 improves test-time scaling of multimodal retrieval.} We compare performance $w.r.t.$ budget of averaged number of tokens per query. Model sizes are reflected in the marker sizes. We consider 2B and 7B embedder sizes. MM Ranker refers to using MLLM (Qwen2.5 VL~\citep{qwen2.5vl} 7B, 32B and 72B) as a computational-expensive baseline, to directly rank \emph{all} the candidates in the retrieval pool. \ours \ecrr refers to our method that reranks the candidates retrieved by the embedder. \ours \ecrr + \qar adds an additional query-target joint reasoning step to introduce more discriminative information for reranking.} %, serving as a theoretical upperbound for computation and accuracy
    \vspace{-10pt}
    \label{fig:placeholder}
\end{wrapfigure}

Multimodal dense retrieval aims to compare heterogeneous modalities within a shared embedding space. Early bi-encoder models~\citep{clip, siglip, coca} focused on modality-invariant representations that capture general semantic similarity between visual and textual content. However, as retrieval scenarios grow more diverse and instruction-driven, the embedding must not only reflect the visual content itself but also how the user specifies the retrieval intent. For example, given an image of a busy street, the instruction “retrieve images with similar traffic patterns” induces a different representation than “retrieve images containing the same storefronts”, despite the visual content being identical. This shift motivates the emerging paradigm of Universal Multimodal Retrieval (UMR)~\citep{vlm2vec, vlm2vecv2}, where a single model must support many retrieval tasks, follow natural-language instructions, and robustly handle heterogeneous input formats.

Multimodal Large Language Models (MLLMs) have become strong candidates for UMR due to their instruction-following ability and unified architecture for various modalities. Existing MLLM-based approaches improve retrieval through architectural designs~\citep{nv_embed}, hard negative mining~\citep{llave, b3, qqmm}, additional synthetic training data~\citep{mmE5}, and improved training procedure with continual contrastive pretraining or auxiliary objectives~\citep{moca, unime}. Recently, Think-Then-Embed~\citep{tte} (\tte) introduces a reasoning step prior to embedding, showing that conditioning the embedder on \emph{Embedding-Centric Reasoning} (ECR) generated by a reasoning model leads to better retrieval performance. 

\begin{wrapfigure}{r}{0.53\textwidth}
    \centering
    \vspace{-12pt}
    \includegraphics[width=\linewidth]{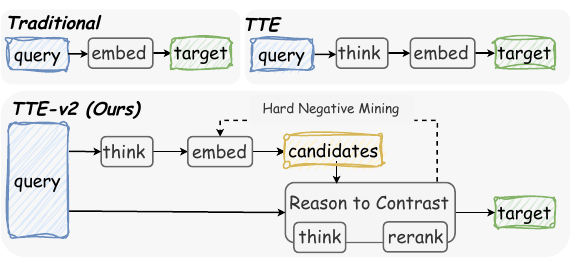}
    \caption{\footnotesize Comparison between different embedding and retrieval frameworks. \textbf{Top left:} traditional \textit{embed-then-retrieve}. \textbf{Top right:} recently proposed \textit{think-then-embed} (\tte)~\citep{tte} procedure. \textbf{Bottom:}  the \textit{reason-to-contrast} framework (\ours) proposed in this work.}
    \vspace{-10pt}
    \label{fig:diagram}
\end{wrapfigure}

% While \tte focuses mainly on the integration of reasoning and embedding, in this work, we instead explore the setting where a  MLLM is leveraged as reranker. To this end, \tte essentially allows flexible integration of more powerful MLLMs at test time, which are too large to be directly trained and served as embedding models. Compared with traditional embedding-model scaling, which typically enhances performance by increasing embedding dimensionality (e.g., Matryoshka Representation Learning) or scaling encoder parameters, in this work, we consider a complementary axis of test-time scaling, through the view of reasoning token budget. 

\tte demonstrates that conditioning an embedder on generated reasoning tokens substantially improves retrieval quality, particularly for complex tasks requiring spatial or semantic interpretation. Importantly, this framework also enables test-time integration of more powerful MLLMs as reasoners, even if these models are too large or costly to train directly as embedding models. Compared with traditional embedding-model scaling--which typically enhances performance by increasing embedding dimensionality (e.g., Matryoshka Representation Learning) or scaling encoder parameters--\tte opens a complementary axis of test-time scaling through reasoning token budget. 

Building on top of \tte, we take a further step toward unlocking the full potential of test-time scaling with extra reasoning tokens. While \tte demonstrates that a short chain of reasoning can substantially enhance the embedder for retrieval, its design restricts the reasoning process to single-side information—the query or the candidate independently—mirroring the inherent limitations of traditional bi-encoders. As a result, the reasoner lacks the ability to form a joint understanding of query–candidate pairs, and discriminative information across candidate lists, where more nuanced joint reasoning could significantly improve ranking quality. To address these limitations, we introduce \textbf{Reason-to-Contrast}, a cascaded multimodal retrieval–reranking framework that enhance query-candidate interaction through joint reasoning and reranking.

\begin{figure*}
    \centering
    \includegraphics[width=\linewidth]{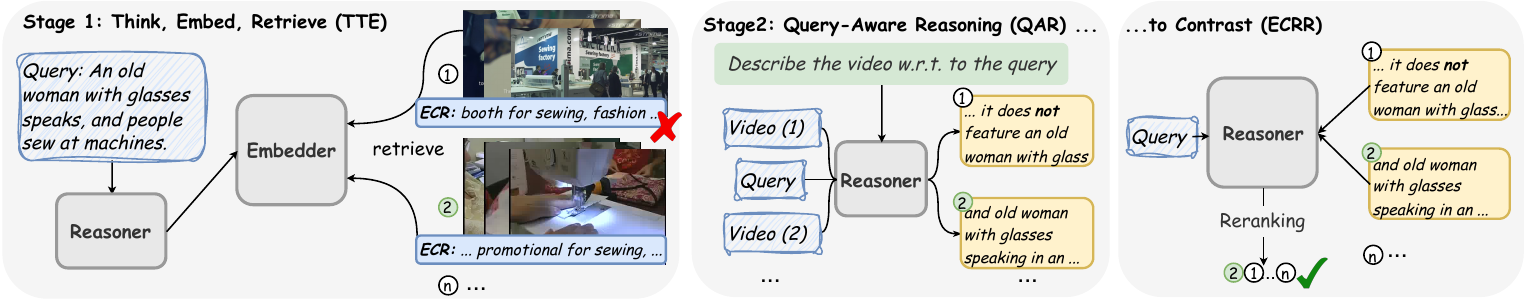}
    \caption{\footnotesize Overall pipeline of our Reason-to-Contrast framework (\ours). In stage 1, we follow the TTE framework: given a query, we retrieve top-$n$ target videos and their ECRs. In Stage 2, we perform the proposed \textit{Reason-to-Contrast} framework, which consists of Query-Aware Reasoning (\qar) stage on the retrieved candidates to incorporate more discriminative information, followed by a ECR-based reranking stage (\ecrr).}
    \label{fig:pipeline}
\end{figure*}

\textbf{Why Reason-to-Contrast?} Multimodal retrievers typically rely on a fast encoder to retrieve an initial top-k candidate set independently. We observe that for tasks with high information density, the coarse embeddings produced by encoders—even when enriched through pre-embedding reasoning as in \tte—are often insufficient to capture the fine-grained distinctions needed for accurate ranking. An example is video retrieval, where the target video contains significantly more information than the query caption. As shown in Figure~\ref{fig:pipeline} and Figure~\ref{fig:qar_sample}, the target ECR shares little relation to the query caption as the query caption refers only to a tiny fraction of the target video. The proposed reason-to-constrast framework solves the challenge by performing reasoning not only for embedding (as in \tte), but also for reranking. In retrieval, reasoning tokens enrich query and candidate embeddings independently. In reranking, they enable explicit joint reasoning between query-candidate pairs and across multiple candidates.

We explore two complementary variants: ECR-based Reranking (\ecrr) and Query-Aware Reasoning (\qar):

\noindent \textbf{(1) ECR-based Reranking (\ecrr).} In this setting, we reuse the ECRs already generated during \tte forward to perform reranking. This design requires no additional reasoning calls, while converting expensive multimodal reranking to lightweight and efficient text-only reranking, thus offering substantial inference cost savings.

\noindent \textbf{(2) Query-Aware Reasoning (\qar).} We further scale up reasoning token budgets by utilizing the MLLM reasoner to perform joint reasoning between the query and retrieved candidates. Instead of treating the query and candidate independently, \qar generates reasoning tokens for candidates that explicitly consider query, thereby adding more discriminative information, which are are then used for reranking as in \ecrr. This can be viewed as a step toward a coarse-to-fine retrieval pipeline, where initial point-wise reasoning serves broad semantic alignment while pair-wise reasoning captures detailed contrastive cues for the final reranking.

\textbf{Hard Negative Mining with Reranking Feedback (\hnm).} While reranking is commonly treated as the final stage of a retrieval pipeline—reshaping the output distribution of a top-k set—we extend its role by introducing a feedback loop from the reranker back to the embedder. This is motivated by \textbf{(1)} ECRs are generated by zero-shot and stronger MLLM, which serves as a stronger and unbiased teacher for Hard Negative Mining (HNM), compared to using the embedder trained on the same dataset as in existing works~\citep{nv_embed, b3, nv_retriever}, and \textbf{(2)} reranker produces the matching score by jointly considering query and candidate, offering far richer semantic and relational understanding than the underlying embedder. As a result, its judgments can be used to mine hard negatives and filter false negatives that are overlooked or misinterpreted by the retriever. By distilling the reranker's distribution through contrastive training, we are able to further improve the retriever on tasks where its own top-$1$ prediction is already sufficiently accurate.

\noindent Our contributions are three folds:
\begin{itemize}
    \item We introduce \ours, a cascaded multimodal retrieval framework that shows the potential for token-wise scaling at test time. Through our proposed \qar and \ecrr, \ours provides over 10\% absolute improvement in zero-shot retrieval compared to the strong \tte baseline.
    \item We introduce, \hnm, a reranker-based hard negative mining and false negative filtering strategy. \hnm provides +1\% overall on MMEB-V2 and outperforms existing embedder-based HNM methods.

    \item \ours establishes a new state of the art on the MMEB-V2 benchmark with an overall score of $75.7\%$, outperforming proprietary models trained on substantially larger private datasets. In video retrieval tasks, \ours improves over the previous best by $6\%$, and notably, even \ours-2B surpasses previous 7B models that rely on extensive external training data.

\end{itemize}

\section{Related Work}
\label{sec:related_work}

\noindent\textbf{Universal Multimodal Retrieval}  Universal Multimodal Retrieval (UMR) seek to unify diverse retrieval modalities and tasks under a single representation space. Recent benchmarks, such as MMEB~\cite{vlm2vec} and MMEB-V2~\citep{vlm2vecv2}, evaluate models across dozens of heterogeneous and unconventional tasks including visual question answering, classification, grounding, document retrieval etc. These tasks require the embedding model to comprehend the task instruction in order to produce the desired embedding specified by the instruction. For instance, in grounding the query could contain an image and a text instruction: \textit{``car second-closest to the camera''}. The embedding model has to first comprehend the instruction, and then produce an embedding as specified by the instruction.

\noindent\textbf{Multimodal Embedding Models} Earlier progress in multimodal embedding typically involves bi-encoder models, where each modality owns separate models~\citep{clip, siglip, flip}. Recent studies show that Multimodal Large Language Models (MLLMs) can achieve state-of-the-art embedding performance after contrastive training, for both text~\cite{nv_embed} and multimodal~\cite{mm_embed, vlm2vec, b3, unime, tte}. These methods typically focus on hard negative mining~\citep{llave, qqmm, b3}, training strategies~\citep{nv_embed, moca} and data curation~\citep{mmE5}. Another line of works~\citep{tte} explore generative capacity of MLLMs for enhancing UMR performance, by introducing intermediate reasoning steps before producing the final embedding.

% \begin{figure*}
%   \centering
%   \begin{subfigure}[t]{0.36\textwidth}
%     \centering
%     \includegraphics[width=\linewidth]{fig/tte_v2_diagram.pdf}
%     \caption{\footnotesize Comparison between different embedding and retrieval frameworks. \textbf{Top left:} traditional \textit{embed-then-retrieve}. \textbf{Top right:} recently proposed \textit{think-then-embed} (TTE)~\citep{tte} procedure. \textbf{Bottom:}  the \textit{reason-to-contrast} framework (\ours) proposed in this work.}
%     \label{fig:diagram}
%   \end{subfigure}
%   \hfill
%   \begin{subfigure}[t]{0.63\textwidth}
%     \centering
%     \includegraphics[width=\linewidth]{fig/tte_v2_demonstration.pdf}
%     \caption{\footnotesize Overall pipeline of our proposed Reason-to-Contrast framework (\ours). In stage 1, we follow the TTE framework: given a query, we retrieve top-$n$ target videos and their ECRs generated through TTE Reasoner. In Stage 2, we perform the proposed \textit{Reason-to-Contrast} framework, which consists of Query-Aware Reasoning (\qar) stage on the retrieved candidates to incorporate more discriminative information, followed by a ECR-based reranking stage (\ecrr).}
%     \label{fig:pipeline}
%   \end{subfigure}
%   \label{fig:embed_head}
% \end{figure*}

\noindent\textbf{Cascaded Multimodal Retrieval and Re-ranking} Cascaded retrieval-then-rerank pipelines are well-studied in information retrieval and recommendation systems~\citep{liu2022neuralrerankingmultistagerecommender}. A common pattern is to use a fast encoder to retrieve a broad set of candidates, followed by a slower or more expressive reranker to refine the top few. Earlier approaches for multimodal reranking typically involve complex modality fusion between query and target~\cite{tan2022instancelevelimageretrievalusing, liu2024candidatesetrerankingcomposed}. Recently studies~\citep{mm_embed} consider MLLMs as zero-shot reranker by prompting the MLLMs to determine whether the candidate match the given query by answering ``True'' or ``False'', and extract the token probability of ``True'' as the reranking score. These methods typically perform multimodal reranking on the multimodal space, which can be computationally expensive and can lead to degraded performance when query and candidate involves complex query and multimodal inputs. On the contrary, \ecrr relies on pre-generated ECR to perform text-only reranking, which is significantly more lightweight, while achieving better performance.

\noindent\textbf{Hard Negative Mining} Contrastive learning benefits from informative negative samples, yet major contrastive training datasets  do not include annotated hard negative samples. Therefore, various approaches are developed to automatically mine hard negatives~\cite{nv_retriever, sfr_embedding, nv_embed, gemini_embedding}. These methods usually utilize a pretrained embedding model to compute embeddings for training samples, and select hard negatives from top-$k$ nearest neighbors. To filter false negative samples, methods like C-Pack~\citep{xiao2024cpackpackedresourcesgeneral} and SFR-Embedding~\citep{sfr_embedding} select hard negatives by excluding top-$n$ similar samples. Other approaches such as NV-Embed~\citep{nv_embed} and Arctic-Embed~\citep{merrick2024arcticembedscalableefficientaccurate} defines a confidence threshold, and the samples with similarity score higher than the threshold are filtered out. These methods all rely on manually defined fixed heuristic to filter false negatives, which may not be optimal under the UMR setting, due to the heterogeneity of training tasks. Recently, B3~\citep{b3} is proposed to construct hard negative batches instead of designated hard negative samples for each training data. However, the effectiveness of this approach diminishes as the batch size scales up. On the contrary, \ecrr utilizes ECR-based reranking to mine hard negative and filter false positives, which is adaptive to tasks and individual positives, while offering complementary information from the views of ECRs and reranker.

\section{Preliminary}

\subsection{MLLM-Based Universal Multimodal Retrieval}

We begin by formalizing the MLLM setting used for UMR. 
Each query $q$ and candidate item $t$ consist of a visual and/or textual input $\mathcal{V}$, and a task-specific instruction $[\texttt{Ins}]$. Both $q$ and $t$ are independently passed through the same MLLM encoder $f_{\theta}$. Let the encoder output a sequence of hidden states; a pooling operator is then applied to obtain the final embedding:
\[    \mathbf{h}_q = \texttt{Pool}\big(f_{\theta}(\mathcal{V}_q, [\texttt{Ins}_q], \mathcal{T}_q)\big), 
    \mathbf{h}_t = \texttt{Pool}\big(f_{\theta}(\mathcal{V}_t, [\texttt{Ins}_t], \mathcal{T}_t)\big).\]
Following prior works~\citep{vlm2vec}, we adopt the final token's hidden state as the pooled representation.

Training is performed using a contrastive objective.  
Given a batch of $N$ matched query--target pairs $(q^i, t^i)$, we optimize a uni-directional InfoNCE loss in which each $q^i$ treats $t^i$ as its positive sample:
\[ \mathcal{L} = -\frac{1}{N} \sum_{i=1}^{N} \log \frac{\phi(\mathbf{h}_{q}^{i}, \mathbf{h}_{t}^{i})}{\displaystyle\sum_{k=1}^{K}  \frac{w_k}{K} \cdot \phi(\mathbf{h}_{q}^{i}, \mathbf{h}_{t}^{k}) +  \displaystyle\sum_{j=1}^{N} \phi(\mathbf{h}_{q}^{i}, \mathbf{h}_{t}^{j})}, \quad
\phi(\mathbf{h}_q, \mathbf{h}_t) = \exp\!\left(\frac{1}{\tau}\cos(\mathbf{h}_q, \mathbf{h}_t)\right), \]
where $\cos(\cdot,\cdot)$ denotes cosine similarity; $\tau$ is a temperature hyper-parameter; $K$ denotes the number of extra hard negatives for each training sample, with a weight $w_k$.

\subsection{Think Then Embed}

Think-then-Embed (TTE)~\cite{tte} introduces an additional reasoning step before producing the final embedding. The reasoning step generates Embedding-Centric-Reasoning (ECR), which are some form of Chain-of-Thought (CoT) ~\citep{cot} that aims at bridging asymmetric gaps between query and target representation. In TTE and this work, ECRs are manually defined for each task, in the form of $\texttt{<think>} \cdots \texttt{</think> Summary}$. The ECR for the text instruction can be a detailed description of the referred object or some regular CoT reasoning. Please refer to the original TTE~\citep{tte} paper for more details. Therefore, each input in TTE is essentially a quadruplet in the form of $\langle \mathcal{V}, \mathcal{T}, [\texttt{Ins}], \psi \rangle$, where $\psi$ denotes the ECR. Figure~\ref{fig:pipeline} (left most) shows the overall framework of TTE. %(\textit{``vehicle second closest to camera''})

\section{Reason-to-Contrast}

\begin{figure*}
    \centering
    \includegraphics[width=\linewidth]{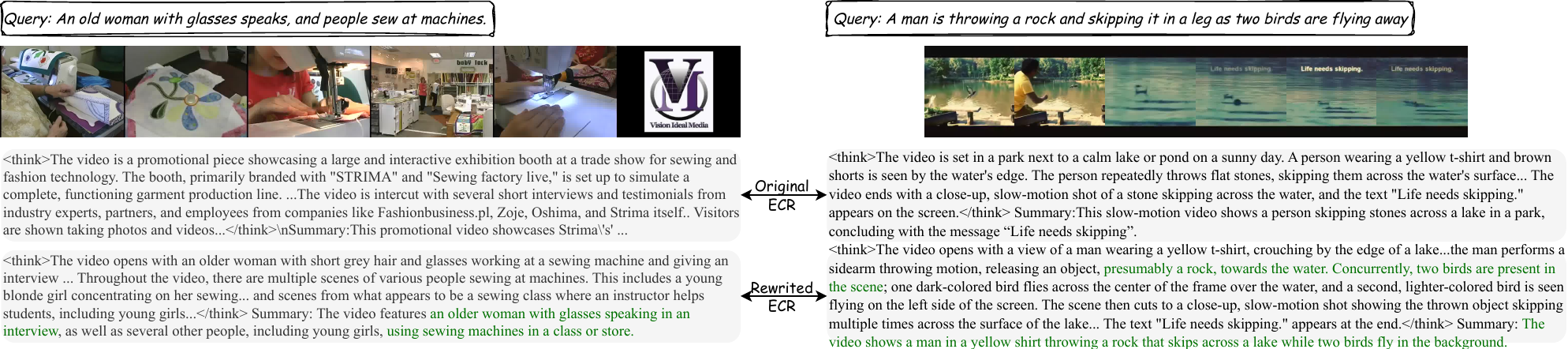}
    \caption{\footnotesize Comparison between the original ECR of a video and the ECR after Query-Aware Rewriting (QAR), from the VATEX T2V test set.}
    \vspace{-10pt}
    \label{fig:qar_sample}
\end{figure*}

\label{sec:method}

In this section we introduce our proposed \ours framework for cascaded multimodal retrieval. \ours involves three components: (1) ECR-based Reranking (Sec.~\ref{sec:ecrr}), (2) Query-Aware ECR Rewriting (Sec.~\ref{sec:qar}), and (3) hard negative mining with ECR Reranking (Sec.~\ref{sec:hn}). Figure~\ref{fig:pipeline} provides an overview of the inference procedure of \ours.

\subsection{ECR-based Reranking}
\label{sec:ecrr}

\begin{wrapfigure}{r}{0.5\textwidth}
    \centering
    \includegraphics[width=\linewidth]{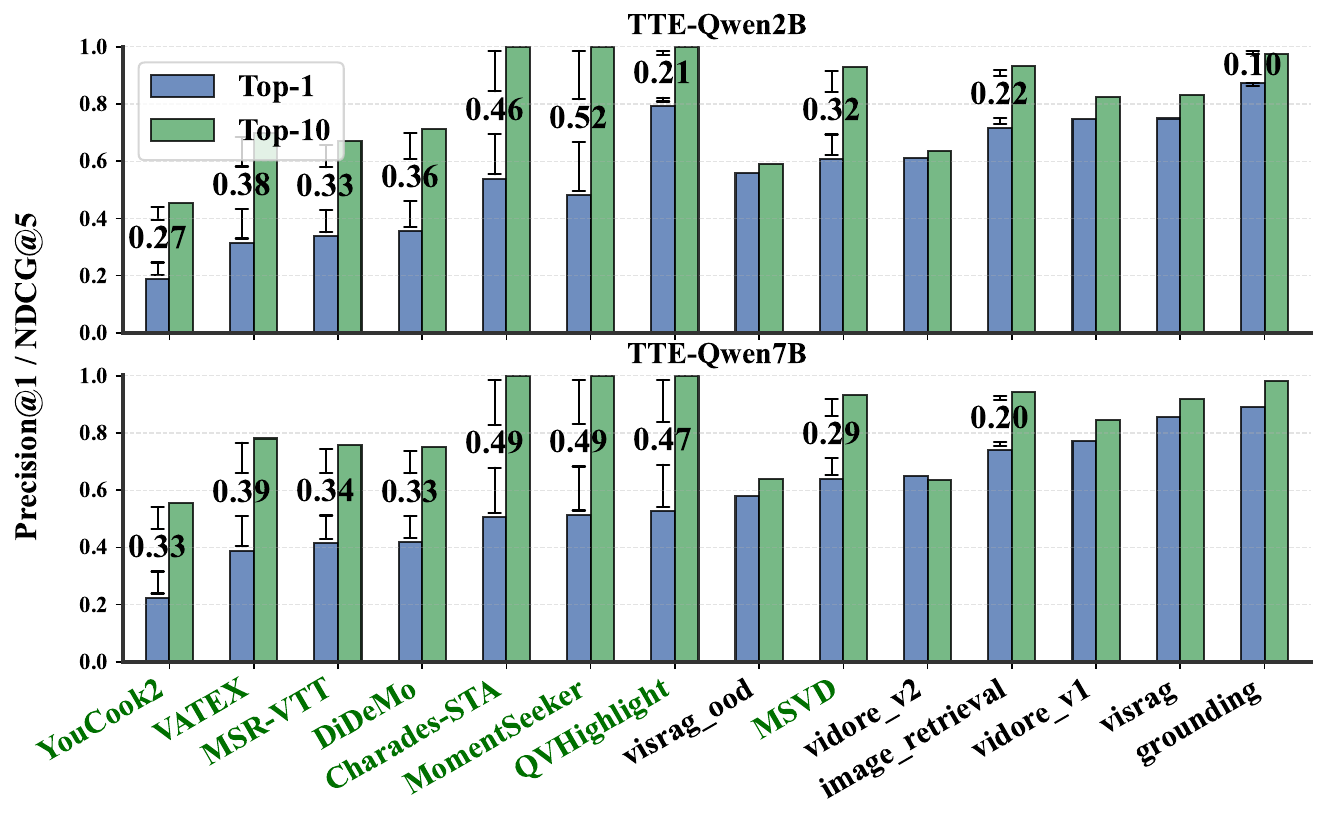}
    \vspace{-20pt}
    \caption{\footnotesize Comparison between top-$1$ and top-$10$ on retrieval tasks in MMEB V2 for TTE-2B (top) and 7B (bottom). For image and video, we show precision, and NDCG for visdoc. We separately show video retrieval tasks (colored in \textcolor[HTML]{007000}{dark green}), while other tasks are averaged by task types.}
    \vspace{-10pt}
    \label{fig:topk_diff}
\end{wrapfigure}

%While TTE has achieved strong retrieval performance with the help of ECRs, it still struggles to solve more challenging tasks, especially for various video retrieval tasks within MMEB V2 evaluation set. As shown in Figure~\ref{fig:topk_diff}, all video retrieval tasks, such as YouCook2, VATEX, and QVHiglight, exhibit a severe performance gap between top-$1$ and top-$k$. 

While TTE has achieved strong retrieval performance with the help of ECRs, it still faces challenges when addressing more complex tasks, particularly various video retrieval tasks within the MMEB V2 evaluation set. As illustrated in Figure~\ref{fig:topk_diff}, many video retrieval tasks exhibit a significant performance gap between top-$1$ and top-$k$ retrieval. This suggests that, although TTE is able to identify relevant candidates within the top-$k$ results, it often struggles to consistently rank the most relevant item at the very top. These observations highlight the increased difficulty of video retrieval tasks compared to other retrieval scenarios. This is more obvious with TTE-2B compared to TTE-7B. We conjecture this is due to videos contain significantly more information than other modalities such as images or text and the embedding model may not be able to encode the details required to perform fine-grained ranking. % within the top-$10$ retrieved candidates, resulting in larger gaps between top-$1$ and top-$10$ accuracy. %(1) Video tasks in MMEB V2 are generally out-of-domain~\footnote{Under the setting of MMEB V2, the training data for video only contains long-form video QA and long-form text(video)-to-video(text), whereas the evaluation tasks are all short VQA and short text-to-video.}, and (2)

To overcome this limitation, we introduce a reranking stage on top of TTE framework to perform zero-shot, text-only reranking. Specifically, under TTE, each multimodal embedding is associated with an Embedding-Centric Reasoning (ECR) trace $\psi$. While in the case of TTE, such ECR is only used to generate the embedding, we propose to reuse the ECRs as reranking criteria. We consider two forms of reranking: pairwise and listwise reranking:

\noindent\textbf{(1) Pairwise reranking,} We adopt an MLLM as a ranking model, which, given a query text and a candidate text, outputs a scalar value $s$ of how well the candidate matches the query. Formally, for the top-$K$ candidates retrieved by query $q$, the reranked candidate order $\mathcal{R}$ is defined as:
\[    \mathcal{R}(q) = \text{argsort}_{k=1\dots K}(s_k), \quad s_k =\mathrm{Reranker}(q, \psi_k )\]
% text-only LLM-based reranking model (e.g., a Qwen3-based reranker)
\noindent\textbf{(2) Listwise reranking.} Given that the ECRs are all text, we can also leverage the LLM for contrasting through reasoning over candidates, i.e., listwise reranking, where the query and top-$k$ candidates' ECRs are passed to a LLM together. We then prompt the LLM to perform zero-shot reranking and directly output the reranked list. Detailed prompts can be found in Appendix A.

%Through extensive experiments on multiple retrieval benchmarks, we found such can achieve significant improvement ($>10\%$). 

\subsection{Query-Aware ECR}

\label{sec:qar}

While we observe notable overall performance gain by using original candidate ECRs, we notice in many cases the ECRs may be over-general and do not contain enough discriminative information relevant to the query. For example, in the left example of Figure~\ref{fig:qar_sample}, the query text is \textit{``An old woman with glasses speaks, and people sew at machines.''}, and the video contains a number of different scenes. The original ECR, while being accurate at the video content, does not contain details that can support or refute the query, as the caption only represents a fraction of the video. 

In order to incorporate more detailed and discriminative information in the candidate ECR \textit{w.r.t} the query, we propose Query-Aware Rewriting. Instead of directly using ECRs as reranking criteria, we first leverage a MLLM to perform pairwise rewrite given the query and candidate input. The rewrited ECRs are then utilized for reranking. This allows the reranker to distinguish hard negatives more accurately. For instance, in the same example in Figure~\ref{fig:qar_sample}, the ECR after QAR contains \textit{``an older woman with glasses speaking...''}, which directly matches with the query caption.

\subsection{Hard Negative Mining with ECR Reranking}
\label{sec:hn}

\begin{wrapfigure}[16]{r}{0.5\columnwidth} % [16]=approx wrapped lines; tweak 12-20
\vspace{-0.8\baselineskip} % optional: small pull-up; remove if it causes trouble
\begin{minipage}{0.48\columnwidth}
\captionsetup{type=algorithm}
\hrule
\vspace{3pt}
\caption{\small ECR-Based Hard Negative Mining}
\label{alg:ecr_hard_neg}
\vspace{-7pt}
\hrule
\vspace{3pt}

\small
\begin{algorithmic}[1]
\Require Training sample $(q, t^+)$; pretrained embedder $f_\theta$; reranker $\mathrm{Reranker}$; retrieval pool size $M$; \# of hard negatives $K$; false negative filtering threshold $\alpha\!\in\!(0,1)$
\Ensure Hard negatives $\mathcal{H} \subseteq \{(t, s)\}$
\State $\{(t_m, \psi_m)\}_{m=1}^M \gets \mathrm{TopM}_{f_\theta}(q)$ \Comment{coarse retrieval}
\State $s^+ \gets \mathrm{Reranker}(q, \psi^+)$;\quad $s_m \gets \mathrm{Reranker}(q, \psi_m)\ \ \forall m$
\State $\pi \gets \mathrm{argsort\_desc}\big(\{s_m\}_{m=1}^M\big)$ \Comment{indices, high $\to$ low}
\State $\mathcal{H} \gets \emptyset$;\quad $j \gets 1$
\While{$|\mathcal{H}| < K$ \textbf{ and } $j \le M$}
  \State $m \gets \pi[j]$
  \If{$s_m < \alpha \cdot s^+$}
    \State $\mathcal{H} \gets \mathcal{H} \cup \{(t_m, s_m)\}$ \Comment{keep as (target, score)}
  \EndIf
  \State $j \gets j + 1$
\EndWhile
\State \Return $\mathcal{H}$
\end{algorithmic}

\vspace{3pt}
\hrule

\end{minipage}
% \vspace{-0.8\baselineskip} % optional: tighten bottom
\vspace{5pt}
\end{wrapfigure}

Another performance bottleneck is lack of informative negative supervision for training. Fortunately, during our \qar and \ecrr, some of the most challenging candidates have been filtered and ranked. By leveraging \ecrr, we propose \hnm, a mechanism for effective hard negative mining.

%While \qar and \ecrr primarily address out-of-domain settings—where coarse embeddings struggle to distinguish among top-$k$ candidates—we also observe that some MMEB-V2 tasks already achieve high top-$1$ accuracy without reranking. For these in-domain cases, the main performance bottleneck is often the lack of informative negative supervision during training. To address this, we propose \hnm, which leverages ECR-based reranking as a mechanism for effective hard negative mining.

The intuition is straightforward: \textbf{(1)} the ECRs used as reranking criteria are generated from more powerful teacher MLLM (e.g., Qwen2.5-VL-72B), which present a potentially more accurate and unbiased view of the input, as compared to using the embedding from the embedder trained on the same dataset. \textbf{(2)} Based on these high-quality ECRs, the reranker considers the query and target \emph{jointly} to produce a similarity score. In fact, \hnm can be seen as an efficient method for HNM with MLLM-as-a-judge, which provides more fine-grained signals for hard negative mining.

% In contrast, prior work in text-only retrieval (e.g., NV-Retriever~\citep{nv_retriever}) mitigates embedder bias by mixing multiple pretrained text models. However, this approach is difficult to extend to UMR: MMEB-V2 includes heterogeneous tasks such as VQA, grounding, document QA, and video retrieval, for which publicly available multimodal embedding models vary widely in domain coverage and performance. Applying alternative multimodal encoders directly to MMEB-V2 leads to substantial degradation, making them ineffective as teachers.

Our \hnm approach is summarized in Algorithm~\ref{alg:ecr_hard_neg}. We adopt a two-stage coarse-to-fine procedure:

\noindent\textbf{(1)} A pretrained embedder performs coarse retrieval to obtain a pool of candidate negatives for each training query.

\noindent\textbf{(2)} ECR-based reranking is then applied to this pool. Candidates ranked above the ground-truth target are treated as false negatives and removed. Among the remaining samples, we use the reranking score $s$ as a hardness measure—either to select the hardest negatives or to assign importance weights $w_k$ in the contrastive loss.

\section{Experiments}

\subsection{Setup}

\noindent\textbf{Datasets.} We evaluate \ours on MMEB-V2~\citep{vlm2vecv2}, a comprehensive benchmark for universal multimodal retrieval spanning image, video, and visual document (visDoc) tasks. Image tasks include classification, VQA, text$\leftrightarrow$image retrieval, and grounding. Video tasks cover VQA, classification, retrieval, and moment retrieval. In total, MMEB-V2 comprises 78 test tasks. Following the evaluation protocol in~\citet{vlm2vecv2}, we report NDCG@5 for visDoc retrieval and Precision@1 for image and video tasks.%
\footnote{As of 11/11/2025, MMEB-V2 contains dataset errors in \textit{ViDoSeek-page} and \textit{MMLongBench-page}. We exclude these datasets for all models. See \url{https://github.com/TIGER-AI-Lab/VLM2Vec/issues/167} for details.}

\noindent\textbf{Models.} For fair comparison with baselines~\citep{vlm2vecv2, tte}, we adopt Qwen2-VL~\citep{qwen2vl} as the base MLLM, with 2B and 7B variants. Our primary baseline is $\text{TTE}_t$, which employs a larger MLLM to generate Embedding-Centric Reasoning (ECR) traces using Qwen2.5-VL-72B. Comparisons include a wide range of dual-encoder baselines~\citep{clip,siglip, magiclens}, and recent MLLM-based embedding methods such as UniME~\citep{unime}, LLaVE~\citep{llave}, and B3~\citep{b3}. %CLIP~\citep{clip}, SigLIP~\citep{siglip}, UniIR~\citep{uniir}, and MagicLens~\citep{magiclens}, as well as 

\noindent\textbf{QAR and ECRR.} We employ Qwen3-Reranker~\citep{qwen3} as a text-only, pairwise reranker and experiment with model sizes of 0.6B, 4B, and 8B. Gemini-2.5-Pro~\citep{gemini2_5} is used for listwise reranking and \qar. We evaluate reranking at top-$K$ candidates with $K \in \{5, 10, 20, 50\}$, focusing primarily on video retrieval tasks within MMEB-V2.

% \noindent\textbf{ECR Reranking and Query-Aware Reranking.} We adopt Qwen3-Reranker~\citep{qwen3} as our text-only, pair-wise reranker, and experiment with model size of 0.6B, 4B and 8B. We use Gemini2.5-Pro~\citep{gemini2_5} to perform listwise reranking and Query-aware pairwise rewriting. We experiment on top-$K$ reranking with $K \in \{5, 10, 20, 50\}$. We experiment ECRR and QAR mainly on the video retrieval tasks under MMEB V2.

% \noindent\textbf{ECR-based hard negative mining.} We use the original VLM2Vec-V2 Qwen2-7B model pretrained on MMEB V2 as the teacher model for generating embeddings to retrieve top-$k$ neighbors for each training query. We use Qwen3-Reranker 8B as the reranking model, with a fixed confidence threshold of 95\% for filtering false negative, according to the reranking score. Following NV\_embed~\citep{nv_embed}, we include 7 hard negative for each training sample. We additionally consider two baselines: (1) random hard negative: we randomly sample negatives from the retrieved top-$K$ candidates. (2) False negative filtering with embedder: we use the same embedder's similarity score as the top-$K$ ranking and filtering criteria. This replicates previous positive-aware hard negative mining approaches~\cite{nv_retriever, nv_embed}.

\noindent\textbf{ECRR-based Hard Negative Mining.} We use Qwen2-7B trained on the same MMEB V2 dataset to retrieve top-$K$ candidates for each query, and Qwen3-Reranker-8B as the reranking model. Following NV-Embed~\citep{nv_embed}, we filter false negatives using a 95\% confidence threshold, and adopt $K=7$ hard negatives per training sample. We consider two baselines: (1) \textit{random negatives}, sampled from the retrieved pool, and (2) \textit{embedder-based filtering}, which relies on similarity scores from the embedder itself, mirroring the positive-aware hard negative mining in~\citep{nv_retriever,nv_embed}.

% \noindent\textbf{Training Details.}
% Following previous work~\cite{tte}, we train with a global batch size of $8192$, a learning rate of $2\times10^{-4}$, and for $2.3$ epochs. The contrastive loss temperature is fixed at $0.02$.
% We apply LoRA~\citep{lora} finetuning with rank $16$ and $\alpha=64$. We use GradCache~\citep{gradcache} to support large per-device batch sizes. We adopt the same dataset-mixing strategy introduced in VLM2Vec-V2: each global batch is interleaved from multiple datasets, and per-dataset sampling weights follow the settings from~\citet{vlm2vecv2}.

\noindent\textbf{Training Details.} We follow the training configuration of TTE~\citep{tte}. Models are trained for 2.3 epochs with a global batch size of $8192$, learning rate of $2\times10^{-4}$, and contrastive temperature $\tau=0.02$. We apply LoRA~\citep{lora} fine-tuning with rank 16 and $\alpha=64$, and utilize GradCache~\citep{gradcache} to enable large per-device batches. Dataset mixing follows VLM2Vec-V2~\citep{vlm2vecv2} with an interleaved batch size of 64.

\subsection{Main Results}

\begin{table*} 
\centering
\renewcommand{\arraystretch}{1.2}
\resizebox{0.8\textwidth}{!}{
\begin{tabular}{lcccccccc}
\toprule
\textbf{Model} & \textbf{MSR-VTT} & \textbf{MSVD} & \textbf{DiDeMo} & \textbf{VATEX} & \textbf{YouCook2} & \textbf{QVHighlight} & \textbf{Charades-STA} & \textbf{Overall} \\
\midrule
\multicolumn{9}{c}{\textbf{Close-sourced Models or w/ Additional Data}} \\
\midrule
IFM-TTE-7B & 52.7 & 73.1 & 49.7 & 51.5 & 31.7 & 64.6 & 50.3 & 53.4 \\
\rowcolor{gray!10}
seed-1.6-embedding & 55.3 & 71.3 & 56.7 & 48.8 & 24.6 & 71.8 & 29.3 & 51.1 \\
RzenEmbed-v2-7B & 50.6 & 72.1 & 57.3 & 47.6 & 27.3 & 64.3 & 24.6 & 49.1 \\
\rowcolor{gray!10}
Ops-MM-embedding-v1-7B & 49.4 & 67.2 & 46.1 & 44.0 & 22.0 & 58.2 & 24.9 & 44.5 \\
RzenEmbed-v1-7B & 45.1 & 64.9 & 41.2 & 38.6 & 19.6 & 60.2 & 22.2 & 41.7 \\
\rowcolor{gray!10}
Ops-MM-embedding-v1-2B & 46.0 & 66.1 & 39.8 & 40.2 & 16.6 & 39.5 & 19.5 & 38.3 \\
RzenEmbed-v1-2B & 43.0 & 58.4 & 39.5 & 33.8 & 16.9 & 51.5 & 18.2 & 37.3 \\
\midrule
\multicolumn{9}{c}{\textbf{Models Trained on MMEB V2 Training Splits}} \\
\midrule
VLM2Vec V2-2B & 28.3 & 48.1 & 30.4 & 26.5 & 10.6 & 49.4 & 20.2 & 30.5 \\
\rowcolor{gray!10}
TTE-2B (Baseline) & 33.9 & 61.7 & 35.7 & 31.5 & 18.7 & 40.3 & 21.4 & 34.7 \\
VLM2Vec V2-7B & 28.3 & 48.1 & 30.4 & 26.5 & 10.6 & 49.4 & 20.2 & 30.5 \\
\rowcolor{gray!10}
UME-R1-7B & 38.9 & 60.8 & 40.0 & 32.7 & 18.5 & 54.9 & 21.9 & 38.2 \\
\tte-7B (Baseline) & 41.9 & 63.9 & 41.8 & 38.9 & 22.5 & 52.7 & 23.0 & 40.7 \\
\midrule
\textbf{(Ours)} \ours-2B & 50.4 & 74.3 & 51.4 & 46.4 & 32.8 & 77.2 & 78.9 & 58.8 \\
\textbf{(Ours)} \ours-7B & \textbf{54.4} & \textbf{72.6} & \textbf{51.1} & \textbf{49.7} & \textbf{33.9} & \textbf{77.2} & \textbf{78.9} & \textbf{59.7} \\
\bottomrule
\end{tabular}
}
\vspace{-5pt}
\caption{\footnotesize Performance comparison on video retrieval tasks from MMEB V2.}
\vspace{-10pt}
\label{tab:video}
\end{table*}

% \midrule
% TTE-2B (Baseline)               & 35.7 & 33.9 & 61.7 & 31.5 & 18.7 & 40.3 & 21.4 & 36.3 \\
% w/ MLLM Reranker               & 37.2 & 35.6 & 58.4 & 29.7 & 20.5 & 15.7 & 12.6 & 36.3 \\
% w/ ECRR (Qwen3-8B)              & 47.9 & 47.7 & 72.2 & 45.2 & 30.1 & 66.2 & 49.3 & 48.6 \\
% w/ ECRR (Gemini)                & 47.1 & 48.3 & 75.6 & 50.4 & 34.8 & 69.4 & 57.1 & 51.2 \\
% w/ QAR + ECRR (Qwen3-8B)        & 51.4 & 50.4 & 74.3 & 46.4 & 32.8 & 77.2 & 78.9 & 51.1 \\
% w/ QAR + ECRR (Gemini)          & 47.9 & 51.6 & 76.4 & 51.8 & 35.5 & 79.2 & 79.7 & 52.6 \\
% \midrule
% TTE-7B (Baseline)               & 41.8 & 41.9 & 63.9 & 38.9 & 22.5 & 52.7 & 23.0 & 41.8 \\
% w/ MLLM Reranker               & 39.5 & 42.3 & 58.6 & 35.6 & 20.3 & 18.4 & 13.5 & 39.3 \\
% w/ ECRR (Qwen3-8B)              & 49.7 & 53.8 & 71.3 & 48.5 & 32.4 & 66.2 & 49.3 & 51.1 \\
% w/ ECRR (Gemini)                & 48.3 & 52.6 & 76.1 & 54.4 & 36.4 & 69.4 & 57.1 & 53.6 \\
% w/ QAR + ECRR (Qwen3-8B)        & 51.1 & 54.4 & 72.6 & 49.7 & 33.9 & 77.2 & 78.9 & 52.3 \\
% w/ QAR + ECRR (Gemini)          & 52.3 & 53.7 & 78.0 & 56.1 & 38.6 & 79.2 & 79.7 & 55.7 \\
% \bottomrule

\noindent\textbf{Video Retrieval.}  
We highlight our main results on video retrieval tasks in Table~\ref{tab:video}. \ours consistently achieves state-of-the-art performance across model scales, surpassing all prior methods by a clear margin. Notably, even the 2B variant exceeds the previous best model by $+5.4\%$.

\begin{wraptable}{r}{0.43\textwidth}
\vspace{-10pt}
\centering
\renewcommand{\arraystretch}{1.2}
\centering
\resizebox{\linewidth}{!}{
% \rowcolors{2}{gray!10}{white}

\begin{tabular}{lcccc}
\toprule
\multirow{2}{*}{\textbf{Model}} & \multicolumn{2}{c}{\textbf{CrossModal-3600}} & \multicolumn{2}{c}{\textbf{XTD10}} \\
\cmidrule(lr){2-3} \cmidrule(lr){4-5}
 & R@1 & R@5 & R@1 & R@5 \\
\midrule
\rowcolor{gray!15}
\multicolumn{5}{c}{\textit{Qwen2-VL 2B}} \\
VLM2Vec-V2 & 19.5 & 36.1 & 33.9 & 57.9 \\
% \rowcolor{gray!10}
$\text{TTE}_t$ & 24.2 & 42.2 & 50.4 & 70.7 \\
\ \ \textit{w/ MLLM Reranker} & 27.5\pls{3.2} & 43.7\pls{1.5} & 48.6\mns{1.8} & 65.4 \mns{5.3} \\
% \rowcolor{gray!10}
\ \ \textit{w/} \ecrr  & 40.2 \pls{16.0} & 49.3 \pls{7.1} & 52.5 \pls{2.1} & 72.4 \pls{1.7} \\
\ \ \textit{w/} \qar + \ecrr & 44.3 \pls{20.1} & 52.7 \pls{10.5} & 55.6 \pls{5.2} & 73.8 \pls{3.1} \\
\midrule
\rowcolor{gray!15}
\multicolumn{5}{c}{\textit{Qwen2-VL 7B}} \\
VLM2Vec-V2 & 31.3 & 52.6 & 42.7 & 66.9 \\
% \rowcolor{gray!10}
$\text{TTE}_t$ & 42.9 & 65.3 & 61.6 & 77.8 \\
\ \ \textit{w/ MLLM Reranker} & 38.6 \mns{4.3} & 58.2 \mns{7.1} & 55.8\mns{5.8} & 71.3\mns{6.5} \\
% \rowcolor{gray!10}
\ \ \textit{w/} \ecrr & 54.1 \pls{11.2} & 70.6 \pls{5.3} & 63.3 \pls{1.7} & 78.6 \pls{0.8} \\
\ \ \textit{w/} \qar + \ecrr & 56.6 \pls{13.7} & 72.5 \pls{7.2} & 64.1 \pls{2.5} & 79.4 \pls{1.6} \\
\bottomrule
\end{tabular}
}
\caption{\footnotesize Effect of Query-Aware ECR (QAR) and ECR-based Reranking (ECRR) on two OOD, text-to-image datasets: CrossModal~\citep{crossmodal3600} and XTD10~\citep{xtd10}. MLLM Zero-Shot refers to the approach~\citep{mm_embed} where a zero-shot MLLM is used as a multimodal reranker.}
\vspace{-15pt}
\label{tab:multilingual}
\end{wraptable}

\noindent\textbf{Multilingual Image Retrieval.}  
Table~\ref{tab:multilingual} reports results on two highly out-of-distribution,  multilingual text-to-image retrieval benchmarks, CrossModal-3600~\citep{crossmodal3600} and XTD10~\citep{xtd10}. As shown in  Table~\ref{tab:multilingual}, both \ecrr and \qar lead to substantial gains over baselines. For example, with the 2B backbone, \ecrr improves Recall@1 on CrossModal-3600 by an absolute $+16.1\%$, highlighting the robustness and generalizability of our approach.

\noindent\textbf{Full MMEB-V2 Benchmark.}  
In Table~\ref{tab:v2_main} we show full results on all 78 MMEB-V2 tasks. \ours-7B attains an overall score of 75.7, outperforming the current state-of-the-art IFM-TTE-7B (75.1). The smaller \ours-2B model also performs strongly, ranking third on the leaderboard. Relative to the $\text{TTE}_t$ baseline, \ours achieves absolute improvements of $+2.7\%$ (2B) and $+3.2\%$ (7B). Beyond video tasks, we further observe consistent 1–2\% gains on image and VisDoc tasks, attributed to our proposed \ecrr-based hard negative mining.

\begin{table*} 
\centering
\renewcommand{\arraystretch}{1.2}
\resizebox{\textwidth}{!}{
\begin{tabular}{l ccccc ccccc ccccc c}
\toprule
\multirow{2}{*}{\textbf{Model}} 
& \multicolumn{5}{c}{\textbf{Image}} 
& \multicolumn{5}{c}{\textbf{Video}} 
& \multicolumn{5}{c}{\textbf{VisDoc}}
& \multirow{2}{*}{\textbf{All}}\\
\cmidrule(lr){2-6} \cmidrule(lr){7-11} \cmidrule(lr){12-16}
& \textbf{CLS} & \textbf{QA} & \textbf{RET} & \textbf{GD} & \textbf{Overall} 
& \textbf{CLS} & \textbf{QA} & \textbf{RET} & \textbf{MRET} & \textbf{Overall} 
& \textbf{VDRv1} & \textbf{VDRv2} & \textbf{VR} & \textbf{OOD} & \textbf{Overall} \\
\midrule
\textbf{\# of Datasets} $\rightarrow$ 
& 10 & 10 & 12 & 4 & 36 
& 5 & 5 & 5 & 3 & 18 
& 10 & 4 & 6 & 2 & 22
& 76
\\
\midrule
\multicolumn{17}{c}{\textbf{Close-sourced Models or w/ Additional Data}} \\
\midrule
IFM-TTE-7B                &
76.7 & 78.5 & 74.6 & 89.3 & 77.9 & 
60.5 & 67.9 & 51.7 & 54.9 & 59.2 &
85.2 & 71.5 & 92.8 & 67.3 & 83.1 & 
75.1
\\
\rowcolor{gray!10}
RzenEmbed-v2-7B           &
70.6 & 71.7 & 78.5 & 92.1 & 75.9 &
58.8 & 63.5 & 51.0 & 45.5 & 55.7 &
89.7 & 60.7 & 88.7 & 69.4 & 82.3 & 
73.0
\\
seed-1.6-embedding        &
76.1 & 74.0 & 77.9 & 91.3 & 77.8 &
55.0 & 60.9 & 51.3 & 53.5 & 55.3 & 
89.5 & 60.8 & 87.9 & 69.4 & 78.4 & 
72.6
\\
\rowcolor{gray!10}
RzenEmbed-v1-7B           &
69.8 & 68.7 & 76.8 & 85.7 & 73.6 &
52.8 & 56.2 & 41.9 & 41.8 & 48.9 &
89.5 & 60.8 & 87.9 & 69.1 & 82.0 &
72.6
\\
Ops-MM-embedding-v1-7B    &
69.7 & 69.6 & 73.1 & 87.2 & 72.7 &
59.7 & 62.2 & 45.7 & 43.2 & 53.8 &
80.1 & 59.6 & 79.3 & 67.1 & 75.0 &
68.9
\\
\rowcolor{gray!10}
RzenEmbed-v1-2B           &
65.3 & 61.7 & 73.8 & 77.9 & 68.5 & 
45.6 & 47.5 & 38.3 & 36.7 & 42.6 &
87.0 & 57.6 & 85.4 & 43.3 & 74.4 &
64.4
\\
Ops-MM-embedding-v1-2B    &
68.1 & 65.1 & 69.2 & 80.9 & 69.0 &
53.6 & 55.7 & 41.8 & 33.7 & 47.6 &
87.0 & 57.6 & 85.4 & 67.9 & 79.5 &
64.6
\\
\rowcolor{gray!10}
interestFM-UIR-CAFe-7B    &
65.2 & 65.6 & 70.0 & 91.2 & 69.8 &
35.8 & 58.7 & 34.4 & 39.5 & 42.4 &
70.7 & 49.6 & 79.5 & 57.0 & 68.0 &
61.7
\\
UniMe-V2-LLaVA-OneVision-7B &
65.6 & 68.7 & 73.1 & 90.8 & 71.8 &
37.2 & 50.6 & 28.9 & 39.6 & 39.0 &
70.7 & 49.6 & 79.5 & 58.9 & 68.1 &
60.7
\\
\rowcolor{gray!10}
GME-7B          & 
57.7 & 34.7 & 71.2 & 59.3 & 56.0 & 
37.4 & 50.4 & 28.4 & 38.2 & 38.6 & 
89.4 & 55.6 & 85.0 & 70.9 & 80.4 & 
58.8      \\
GME-2B             & 
54.4 & 29.9 & 66.9 & 55.5 & 51.9 & 
34.9 & 42.0 & 25.6 & 32.4 & 33.9 & 
86.1 & 54.0 & 82.5 & 69.7 & 77.8 & 
55.1      \\
\rowcolor{gray!10}
ColPali v1.3-3B         & 
40.3 & 11.5 & 48.1 & 40.3 & 34.9 & 
26.7 & 37.8 & 21.6 & 25.5 & 28.2 & 
83.6 & 52.0 & 81.1 & 72.4 & 76.2 & 
45.2
\\
LamRA-Qwen2.5-7B          & 
51.7 & 34.1 & 66.9 & 56.7 & 52.4 & 
32.9 & 42.6 & 23.2 & 37.6 & 33.7 & 
56.3 & 33.3 & 58.2 & 62.0 & 53.1 & 
48.2 \\
% interestFM-UIR-CAFe-7B    &
% 63.6 & 61.7 & 69.1 & 87.6 & 67.6 &
% 35.8 & 58.7 & 34.4 & 39.5 & 42.4 &
% 70.7 & 49.6 & 79.5 & 38.1 & 63.9 &
% 60.6 \\
\midrule
\multicolumn{17}{c}{\textbf{Models Trained on MMEB V2 Training Splits}} \\
\midrule
% VLM2Vec-Qwen2VL (2B)      & 
% 58.7 & 49.3 & 65.0 & 72.9 & 59.7 & 
% 33.4 & 30.5 & 20.6 & 33.0 & 29.0 & 
% 49.8 & 13.5 & 51.8 & 33.5 & 41.6 & 
% 47.0    \\
% B3\_Qwen2\_2B             &
% 67.0 & 61.2 & 70.9 & 79.9 & 68.1 &
% - & - & - & - & - & 
% - & - & - & - & - & 
% -    \\
% LLaVE-2B                &
% 62.1 & 60.2 & 65.2 & 84.9 & 65.2 &
% - & - & - & - & - & 
% - & - & - & - & - & 
% -    \\
% UniME (4.2B)              &
% 54.8 & 55.9 & 64.5 & 81.8 & 64.2 &
% - & - & - & - & - & 
% - & - & - & - & - & 
% -    \\
% UniME (7B)              &
% 60.6 & 52.9 & 67.9 & 85.1 & 66.6 &
% - & - & - & - & - & 
% - & - & - & - & - & 
% -    \\
% LLaVE-7B                &
% 65.7 & 65.4 & 70.9 & 91.9 & 70.3 &
% - & - & - & - & - & 
% - & - & - & - & - & 
% -    \\
% B3\_Qwen2\_7B           &
% 70.0 & 66.5  &74.1 & 84.6 & 72.0 &
% - & - & - & - & - & 
% - & - & - & - & - & 
% -    \\
% \rowcolor{gray!10}
% VLM2Vec-Qwen2VL (7B)      & 
% 62.7 & 56.9 & 69.4 & 82.2 & \textbf{65.5} & 
% 39.1 & 30.0 & 29.0 & 40.6 & 34.0 & 
% 56.9 & 9.4 & 59.1 & 38.1 & 46.4 & 
% 52.3 \\
VLM2Vec-V2-2B             & 
62.9 & 56.3 & 69.5 & 77.3 & 64.9 &
39.3 & 34.3 & 28.8 & 38.5 & 34.9 & 
75.5 & 44.9 & 79.4 & 62.0 & 69.8 & 
59.1
\\
\rowcolor{gray!10}
VLM2Vec-V2-7B     &
65.7 & 61.5 & 70.0 & 85.2 & 68.1 &
45.9 & 33.9 & 27.6 & 39.3 & 36.4 & 
78.8 & 52.6 & 82.7 & 67.2 & 74.1 & 
62.4    \\
UME-R1-7B                 &
67.1 & 69.2 & 71.9 & 84.8 & 71.2 &
48.6 & 60.7 & 38.2 & 39.3 & 47.5 &
75.7 & 50.5 & 83.7 & 33.5 & 68.4 &
64.8
\\
\rowcolor{gray!10}
$\text{TTE}_t$-2B         &
76.6 & 76.8 & 71.5 & 87.2 & 76.1 &
56.1 & 65.3 & 34.1 & 33.8 & 47.9 &
81.1  &62.4 & 84.7 & 65.9 & 77.3 &
70.0
\\
$\text{TTE}_t$-7B&
76.7 & 78.6 & 74.3 & 89.3 & 77.8 &
57.5 & 68.2 & 38.0 & 39.3 & 52.0 &
83.7 & 63.6 & 91.4 & 67.7 & 80.7 &
72.5
\\
\midrule
% \textbf{(Ours)} \ours-2B $(\mathrm{HNM})$  &
% 77.6 & 77.8 & 72.5 & 88.2 & 77.1 &
% 57.7 & 65.7 & 36.4 & 37.0 & 50.5 &
% 82.5 & 63.8 & 83.3 & 47.9 & 77.6 &
% 71.0
% \\
% \rowcolor{gray!10}
\textbf{(Ours)} \ours-2B  &
77.6 & 77.8 & 72.5 & 88.2 & 77.1 &
57.7 & 65.7 & 49.3 & 68.0 & 59.3 &
82.5 & 63.8 & 83.3 & 47.9 & 77.6 &
72.8
\\
% \textbf{(Ours)} \ours-7B $(\mathrm{HNM})$        &
% 78.1 & 79.0 & 76.3 & 91.6 & 79.2 &
% 58.3 & 66.9 & 41.6 & 59.8 & 56.3 &
% 85.4 & 63.3 & 94.3 & 53.9 & 82.3 &
% 74.6
% \\
\textbf{(Ours)} \ours-7B           &
78.1 & 79.0 & 76.3 & 91.6 & 79.2 &
58.3 & 66.9 & 52.3 & 68.0 & 60.7 &
85.4 & 63.3 & 94.3 & 69.0 & 82.3 &
75.7
\\

% InternVideo2 (1B) & 42.5    & 7.3    & 45.5    & 67.9    & 36.6    & \textbf{55.7} & \textbf{38.7} & \textbf{40.8} & 20.2 & \textbf{41.1} & - & -      \\  % reported using older eval code for COLM submission, missing Visdoc tasks and ActivityNetQA
% E5-V (8B)         & 21.8 & 4.9 & 11.5  & 19.0  & 13.3 & 25.9 & 26.9 & 21.4 & 38.5 & 27.0 & -  & -     \\  % reported using older eval code for COLM submission, missing ActivityNetQA

\bottomrule
\end{tabular}}

\caption{\footnotesize Performance comparison on MMEB-V2~\citep{vlm2vecv2}. Task abbrev: CLS (classification), QA (question answering), RET (retrieval), GD (grounding), MRET (moment retrieval), VDR (ViDoRe), VR (VisRAG), and OOD (out-of-domain).}
% \semih{Explain what "-" means with a justification of why}}
\label{tab:v2_main}
\vspace{-10pt}
\end{table*}

\subsection{Ablations}

\noindent\textbf{Effect of Reranking Methods.} In Table~\ref{tab:reranking_abl} we compare with different reranking strategies, including the MLLM-based zero-shot reranker~\citep{mm_embed}, and our proposed \ecrr in pairwise and listwise forms. Both pairwise and listwise \ecrr yield substantial improvements over the baseline \tte, with the listwise variant achieving slightly higher accuracy (typically within 3\%). For example, under the \tte-2B setting, \ecrr-pairwise and \ecrr-listwise improve the baseline (37.2) to 47.9 and 47.1 on DiDeMo, respectively—an absolute gain exceeding 10\%. The improvements are particularly pronounced for the smaller 2B model, effectively narrowing the gap between weaker and stronger embedders (e.g., $4.5\% \rightarrow 1.2\%$ for pairwise \ecrr).

\begin{wraptable}{r}{0.5\textwidth}
\vspace{-10pt}
\centering
\renewcommand{\arraystretch}{1.2}
\resizebox{\linewidth}{!}{
\begin{tabular}{cccccccccc}
\toprule
\ecrr & \qar & \textbf{DiD.} & \textbf{MSR.} & \textbf{MSV.} & \textbf{VAT.} & \textbf{You.} & \textbf{QVH.} & \textbf{Char.} & \textbf{Overall} \\
\midrule
\rowcolor{gray!15}
\multicolumn{10}{c}{\tte-2B} \\
\no         & \no     & 35.7 & 33.9 & 61.7 & 31.5 & 18.7 & 40.3 & 21.4 & 36.3 \\
\rowcolor{gray!10}
MLLM R.   &   \no             & 37.2 & 35.6 & 58.4 & 29.7 & 20.5 & 26.8 & 12.6 & 36.3\pls{0.0} \\
P. & \no              & 47.9 & 47.7 & 72.2 & 39.2 & 30.1 & 66.2 & 49.3 & 48.6\pls{12.3} \\
\rowcolor{gray!10}
L. & \no              & 47.1 & 48.3 & 75.6 & 41.6 & 34.8 & 69.4 & 57.1 & 51.2\pls{14.9} \\
P. & \yes        & 51.4 & 50.4 & 74.3 & 46.4 & 32.8 & 77.2 & 78.9 & 51.1\pls{14.8} \\
\rowcolor{gray!10}
L. & \yes          & 47.9 & 51.6 & 76.4 & 51.8 & 35.5 & 79.2 & 79.7 & 52.6\pls{16.3} \\
\midrule
\rowcolor{gray!15}
\multicolumn{10}{c}{\tte-7B} \\
\no         & \no               & 41.8 & 41.9 & 63.9 & 38.9 & 22.5 & 52.7 & 23.0 & 41.8 \\
\rowcolor{gray!10}
MLLM R.   &   \no                & 39.5 & 42.3 & 58.6 & 35.6 & 20.3 & 30.6 & 13.5 & 39.3\mns{2.5} \\
P. & \no             & 49.7 & 53.8 & 71.3 & 44.5 & 32.4 & 66.2 & 49.3 & 51.1\pls{9.3} \\
\rowcolor{gray!10}
L. & \no                & 48.3 & 52.6 & 76.1 & 46.8 & 36.4 & 69.4 & 57.1 & 53.6\pls{11.8} \\
P. & \yes        & 51.1 & 54.4 & 72.6 & 49.7 & 33.9 & 77.2 & 78.9 & 52.3\pls{10.5} \\
\rowcolor{gray!10}
L. & \yes         & 52.3 & 53.7 & 78.0 & 56.1 & 38.6 & 79.2 & 79.7 & 55.7\pls{13.9} \\
\bottomrule
\end{tabular}
}
\vspace{-5pt}
\caption{\footnotesize Ablations on \ecrr and \qar for video retrieval tasks, with top-$10$ reranking candidates. Shorthands: \textbf{MLLM R.}: MLLM-based multimodal reranker. \textbf{L.}: listwise reranking using Gemini2.5Pro; \textbf{P.}: pairwise reranking using Qwen3 8B. Datasets from left to right: DiDeMo, MSR-VTT, MSVD, VATEX, YouCook2, QVHighlight, Charades-STA.}
\vspace{-15pt}
\label{tab:reranking_abl}
\end{wraptable}

The MLLM-based zero-shot reranker, however often fails to enhance retrieval accuracy and can severely degrade performance, particularly with stronger retrievers (e.g. \tte-7B). This aligns with prior findings from MM\_Embed~\citep{mm_embed}, which observed that such rerankers are primarily effective for VQA-style tasks but less so for retrieval. Notably, the zero-shot reranker causes sharp drops on QVHighlight and Charades-STA ($-14.5\%$ and $-8.8\%$ on \tte-2B), likely due to the MLLM’s limited ability to process multiple video inputs simultaneously.

% \begin{minipage}[t]{0.5\textwidth}
% \centering
% \includegraphics[width=0.32\linewidth]{fig/reranker_MSR-VTT.pdf}
% \includegraphics[width=0.32\linewidth]{fig/reranker_VATEX.pdf}
% \includegraphics[width=0.32\linewidth]{fig/reranker_Charades-STA.pdf}
% \vspace{-6pt}
% \captionof{figure}{\footnotesize Effects of reranker sizes versus top-$k$ candidates in reranking.}
% \label{fig:rerank}
% \end{minipage}

\noindent\textbf{Effect of Query-Aware Reranking (QAR).} 
In Table~\ref{tab:reranking_abl} we can also observe that enabling \qar consistently improves performance across tasks, with the largest gains on datasets requiring complex joint reasoning between query and target—such as QVHighlight and Charades-STA, where \tte-7B improves by more than 10\%. We also observe notable improvements on datasets like VATEX ($>5\%$ absolute gain), where targets contain significantly more information than the query. In such cases, the original ECR may overlook query-relevant cues, whereas \qar refines the reasoning trace to emphasize discriminative content aligned with the query. Figure~\ref{fig:qar_sample} illustrates two examples from VATEX, showing that \qar effectively grounds reasoning to the specific query caption.

\noindent\textbf{Effect of Reranker Size and Top-$k$.} 
Figure~\ref{fig:topk_diff} analyzes the effect of reranker model size and the number of top-$k$ candidates ($k \in \{5, 10, 20, 50\}$). Even a lightweight 0.6B reranker provides clear gains over the \tte baseline, while scaling up to 8B brings further yet moderate improvement. Conversely, increasing $k$ beyond 10 yields limited benefit and can even degrade performance (e.g., 0.6B reranker on VATEX), likely due to additional distractors introduced in larger candidate pools. These results indicate that \ecrr achieves strong performance without requiring extensive reranking depth or model scaling.

\begin{table*} 
\centering
\renewcommand{\arraystretch}{1.2}
\resizebox{\textwidth}{!}{
\begin{tabular}{l ccccc ccccc ccccc c}
\toprule
\multirow{2}{*}{\textbf{Model}} 
& \multicolumn{5}{c}{\textbf{Image}} 
& \multicolumn{5}{c}{\textbf{Video}} 
& \multicolumn{5}{c}{\textbf{VisDoc}}
& \multirow{2}{*}{\textbf{All}}\\
\cmidrule(lr){2-6} \cmidrule(lr){7-11} \cmidrule(lr){12-16}
& \textbf{CLS} & \textbf{QA} & \textbf{RET} & \textbf{GD} & \textbf{Overall} 
& \textbf{CLS} & \textbf{QA} & \textbf{RET} & \textbf{MRET} & \textbf{Overall} 
& \textbf{VDRv1} & \textbf{VDRv2} & \textbf{VR} & \textbf{OOD} & \textbf{Overall} \\
\midrule
\rowcolor{gray!20}
\multicolumn{17}{c}{\tte-2B} \\

Baseline ($\text{TTE}_t$)    & 76.6 & 76.8 & 71.7 & 87.3 & 76.2 & 57.4 & 65.3 & 36.1 & 36.7 & 50.1 & 81.1 & 62.4 & 82.0 & 45.8 & 76.3 & 70.1 \\
\rowcolor{gray!10}
\ \ \textit{w/} B3~\cite{b3} & 76.9 & 77.2 & 72.3 & 87.6 & 76.7 & 57.5 & 64.9 & 36.0 & 35.3 & 49.9 & 81.8 & 63.2 & 82.1 & 45.8 & 76.8 & 70.3 \\

\ \ \textit{w/ Random $\mathrm{HNM}$}    & 76.2 & 76.5 & 71.3 & 87.2 & 75.9 & 57.5 & 65.4 & 35.9 & 36.3 & 50.4 & 80.5 & 62.5 & 82.6 & 45.0 & 76.4 & 70.0 \\
\rowcolor{gray!10}
\ \  \textit{w/ Embedder $\mathrm{HNM}$} & 76.8 & 77.0 & 71.9 & 87.5 & 76.4 & 56.5 & 65.6 & 36.4 & 36.4 & 51.2 & 81.7 & 62.4 & 82.5 & 46.4 & 76.7 & 70.3 \\
\ \  \textit{w/ \hnm}     & 77.3 & 77.6 & 72.5 & 88.3 & 77.0 & 57.7 & 65.7 & 36.5 & 36.4 & 50.5 & 82.2 & 63.5 & 82.9 & 46.7 & 77.1 & 70.7 \\
\rowcolor{gray!10}
\ \  \textit{w/ Weighted \hnm}
                             & 77.6 & 77.8 & 72.5 & 88.2 & 77.1 & 57.3 & 65.9 & 36.4 & 37.0 & 50.5 & 82.5 & 63.8 & 83.3 & 47.9 & 77.6 & 71.0 \\

\midrule
\rowcolor{gray!20}
\multicolumn{17}{c}{\tte-7B} \\
Baseline ($\text{TTE}_t$)    & 76.7 & 78.6 & 74.3 & 89.3 & 77.8 & 58.1 & 68.2 & 41.8 & 60.0 & 56.7 & 83.7 & 63.6 & 91.4 & 67.6 & 80.7 & 73.7 \\
\rowcolor{gray!10}
\ \ \textit{w/} B3~\citep{b3}                & 76.8 & 78.1 & 75.7 & 88.4 & 78.1 & 58.4 & 66.6 & 40.5 & 60.5 & 56.1 & 84.8 & 70.6 & 89.3 & 65.9 & 81.7 & 73.9 \\
\ \ \textit{w/ Random $\mathrm{HNM}$}    & 77.1 & 78.4 & 73.9 & 89.6 & 77.8 & 58.3 & 67.1 & 42.2 & 60.1 & 56.6 & 83.4 & 61.3 & 89.6 & 67.5 & 79.6 & 73.3 \\
\rowcolor{gray!10}
\ \  \textit{w/ Embedder $\mathrm{HNM}$} & 77.2 & 78.3 & 75.2 & 90.2 & 78.3 & 58.6 & 67.4 & 41.8 & 60.4 & 56.6 & 84.6 & 62.4 & 92.4 & 66.4 & 81.1 & 74.0 \\
\ \  \textit{w/} \hnm     & 77.6 & 79.4 & 75.7 & 90.1 & 78.8 & 58.2 & 66.8 & 41.6 & 59.9 & 56.3 & 85.2 & 64.3 & 92.7 & 81.4 & 83.1 & 74.3 \\
\rowcolor{gray!10}
\ \  \textit{w/ Weighted \hnm}
         & 78.1 & 78.8 & 76.1 & 91.6 & 79.1 & 58.3 & 66.9 & 41.6 & 59.8 & 56.3 & 85.4 & 64.0 & 94.3 & 69.0 & 82.5 & 74.7 \\
\bottomrule
\end{tabular}}
\caption{\footnotesize Ablations on hard negative mining strategies on MMEB V2. \textit{Random HN}: random negatives from retrieved candidates; \textit{Embedder $\mathrm{HNM}$}: using a pretrained embedding model to select top-$k$ similar candidates and filter false negatives based on cosine similarity; \hnm: our proposed method, which first use a pretrained embedding model to retrieve coarse candidates, then use \ecrr to select top-$k$ reranked candidates and filter false negatives based on ranking score \textit{w.r.t.} true positive; \textit{Weighted \hnm}: using ranking score to reweigh each hard negative during training.}

\label{tab:hn_abl}
\vspace{-10pt}
\end{table*}

\begin{wrapfigure}{l}{0.5\textwidth}
\vspace{-10pt}
\centering
    \begin{subfigure}[t]{0.16\textwidth}
        \includegraphics[width=\linewidth]{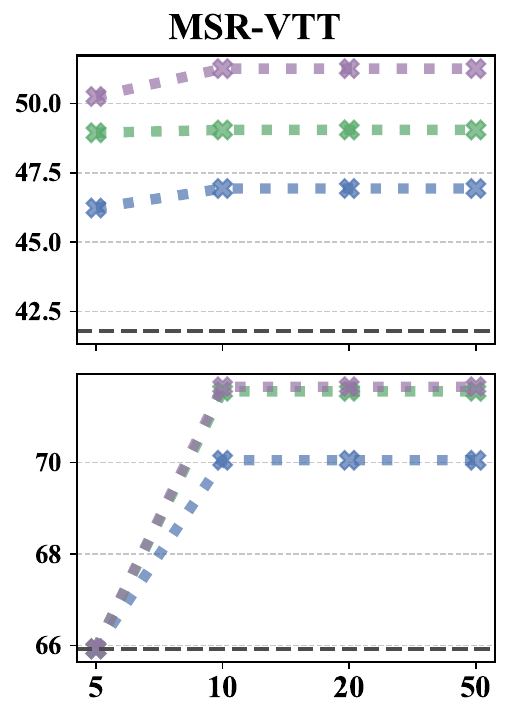}
    \end{subfigure}
    \begin{subfigure}[t]{0.16\textwidth}
        \includegraphics[width=\linewidth]{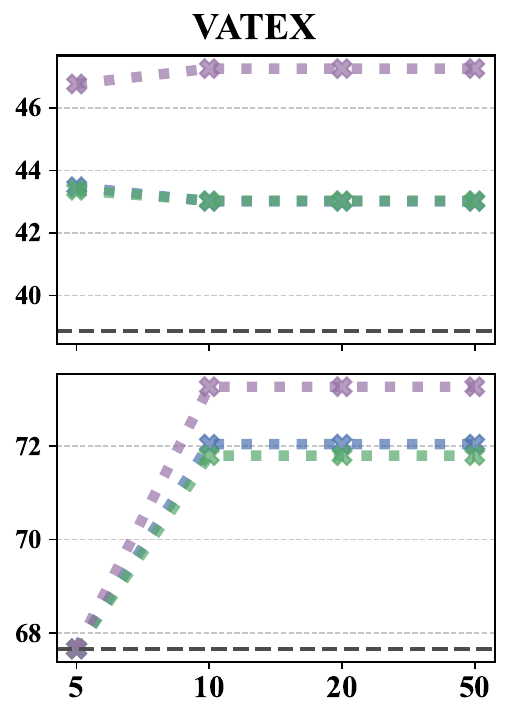}
    \end{subfigure}
    \begin{subfigure}[t]{0.16\textwidth}
        \includegraphics[width=\linewidth]{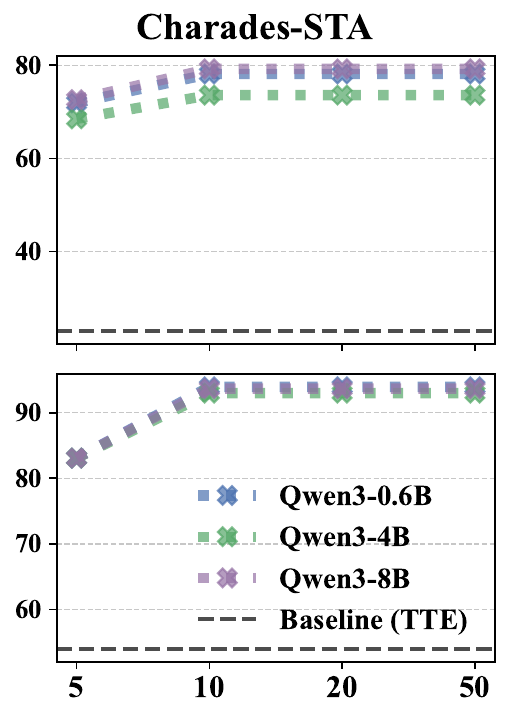}
    \end{subfigure}
    \vspace{-10pt}
    \captionof{figure}{\footnotesize Effects of reranker sizes versus top-$k$ candidates in reranking.}
    \label{fig:rerank}
    % \vspace{-18pt}
\end{wrapfigure}

\noindent\textbf{Effect of \ecrr-based Hard Negative Mining.} 
Table~\ref{tab:hn_abl} presents ablations on our \hnm compared with several hard negative mining baselines. Our method yields the highest overall performance. The random baseline fails to bring consistent improvement and can even harm accuracy due to false negatives. The recent B3~\citep{b3} approach provides marginal gains ($<0.2\%$) on both 2B and 7B backbones, likely because the large batch size (8192) diminishes its batch-construction effect~\citep{b3}. Compared to the embedder-based HNM used in~\citep{nv_retriever}, which achieves 74.0\% with \tte-7B, our \ecrr-based HNM reaches 74.7\%, confirming that \ecrr is able to serve as a stronger criterion for selecting hard negatives and filtering false positives.

\begin{wrapfigure}{r}{0.47\linewidth}
    \centering
    \vspace{-10pt}
    \includegraphics[width=\linewidth]{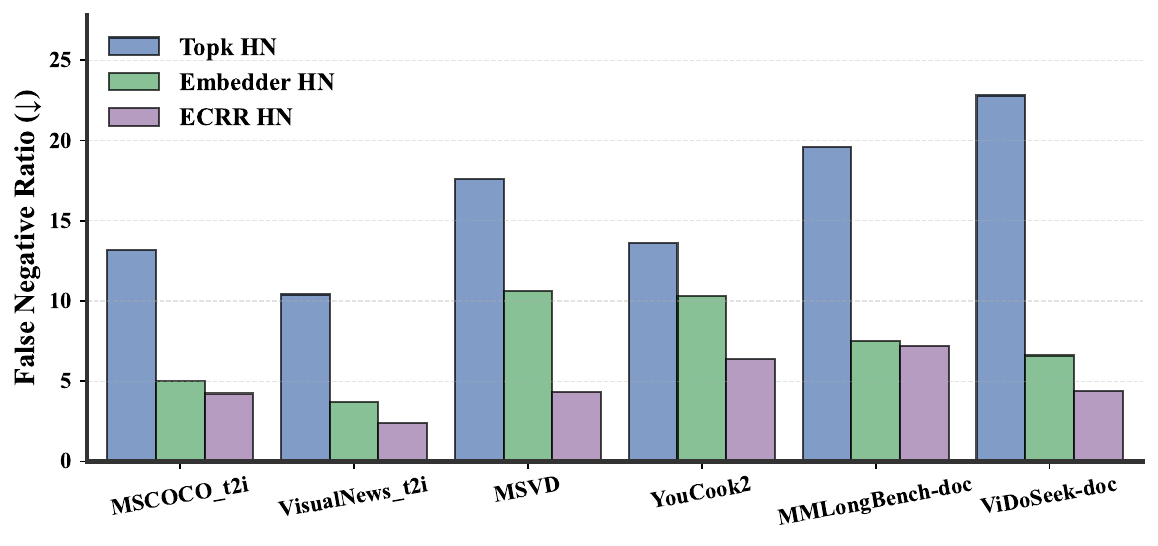}
    \vspace{-25pt}
    \caption{\footnotesize False negative ratio (\%) comparison determined by MLLM-as-a-judge on 6 test datasets in MMEB V2.}
    \label{fig:fn_ratio}
    \vspace{-15pt}
\end{wrapfigure}

Figure~\ref{fig:fn_ratio} reports false negative ratios across six test datasets. Following NV-Retriever~\citep{nv_retriever}, we prompt a large MLLM (Qwen2.5-VL 72B) to determine whether a non-positive candidate is relevant to the query. \ecrr-based HNM consistently reduces false negatives compared to naive top-$k$ selection and embedder-based filtering, supporting our hypothesis that \ecrr offers a more reliable judgment mechanism than pretrained embedders. In particular, we can observe larger gain of reranker (compared to embedder) on the MSVD and YouCook2 datasets, which are more complex retrieval tasks. This is expected, as embedder already performs strong on easier datasets such as MSCOCO\_t2i and VisualNews\_t2i.

% in the form of ``$\langle \textit{visual} \rangle$ \textit{Determine if the provided image(video) is relevant to the query below. Answer yes or no.Query:}''.

% \input{sec/6_ablations}
\section{Conclusion}
We propose \ours, a simple yet effective retrieval framework for test time scaling. We introduce Query-Aware ECR (\qar) and ECR-based Reranking (\ecrr) to improve retrieval in a coarse-to-fine manner. We also propose a \ecrr-based hard negative mining approach for false negative filtering, and to distill the knowledge of \ecrr into the embedding model. \ours achieves SOTA performance on MMEB V2, without relying on additional data.

\clearpage
% \section*{Ethics Statement}
% We reviewed the ICLR Code of Ethics carefully and do not observe potential concerns for our work. 

% \section*{Reproducibility Statement}
% We made our best efforts to comprehensively document the implementation details. Training hyper-parameters and model architectures are discussed in Section~\ref{s_exp_setup} and~\ref{sec:sft_cl}. We include the dataset construction details including all the example prompts we used in Section~\ref{sec:prompts_ecrs}. For evaluation, as mentioned in Section~\ref{s_exp_setup}, we strictly follow the official setup with the codebase released by the original authors. We reveal the full results of each task on the benchmark, without average, in Section~\ref{sec:full_results}.

\bibliography{main}
\bibliographystyle{assets/plainnat}

\newpage
\beginappendix

\section{More ablations}

\paragraph{Effect of ECR quality on reranking.}
Since \ecrr depends directly on the generated ECRs, a key question is how the quality of these ECRs influences reranking performance. Table~\ref{tab:abl_reranking_ecr_model} evaluates this effect by generating ECRs with Multimodal Large Language Models of different capacities (Qwen2.5 VL 32B, 72B~\citep{qwen2.5vl}, and Gemini 2.5 Pro~\citep{gemini2_5}). We then apply TTE-based retrieval~\citep{tte}, which embeds both the original input and the generated ECRs, followed by \ecrr using a fixed Qwen3 8B reranker~\citep{qwen3}. As a reference, we also compare with MLLM-based zero-shot reranking~\citep{mm_embed} at corresponding model sizes.

Overall, \ecrr consistently outperforms MLLM-based zero-shot reranking across all ECR generators. Notably, it achieves strong results even when the ECRs are produced by a weaker MLLM. For example, ECRs generated by Qwen2.5 VL 32B already yield an overall score of 36.8 with \ecrr, compared with 27.0 for zero-shot reranking using Qwen2.5 VL 72B. This gap is largely driven by the poor robustness of zero-shot reranking on QVHighlight and Charades-STA, which require multi-video temporal reasoning. On standard text-to-video retrieval tasks, \ecrr performs competitively with zero-shot reranking at the same ECR model size, while remaining far more efficient since reranking is text-only and uses a lightweight reranker. With ECRs generated by Gemini 2.5 Pro, \ecrr surpasses the strongest zero-shot reranker (Qwen2.5 VL 72B) by a substantial margin on all tasks.

\paragraph{Effectiveness of Query-Aware Reasoning.}
Table~\ref{tab:abl_qar} reports the performance of \qar using different MLLM sizes (Qwen2.5 VL 32B, 72B, and Gemini 2.5 Pro), with ECRs fixed to those generated by Gemini 2.5 Pro. \qar achieves the strongest overall results on video tasks, with marked improvements on datasets such as VATEX, QVHighlight, and Charades-STA. These benchmarks require more fine-grained ranking where the baseline embedding often lacks sufficient discriminative power.

Performance trends also reflect the capability of the model used for \qar. When using Qwen2.5 VL 32B, \qar can underperform relative to \ecrr due to the large gap between this model and Gemini 2.5 Pro. With Qwen2.5 VL 72B, \qar begins to surpass \ecrr, and with Gemini 2.5 Pro it delivers substantial gains. For example, on VATEX, \qar with Gemini 2.5 Pro improves over \ecrr by more than seven percent.

\begin{wrapfigure}{r}{0.5\textwidth}
    \centering
    \begin{subfigure}[t]{0.15\textwidth}
        \includegraphics[width=\linewidth]{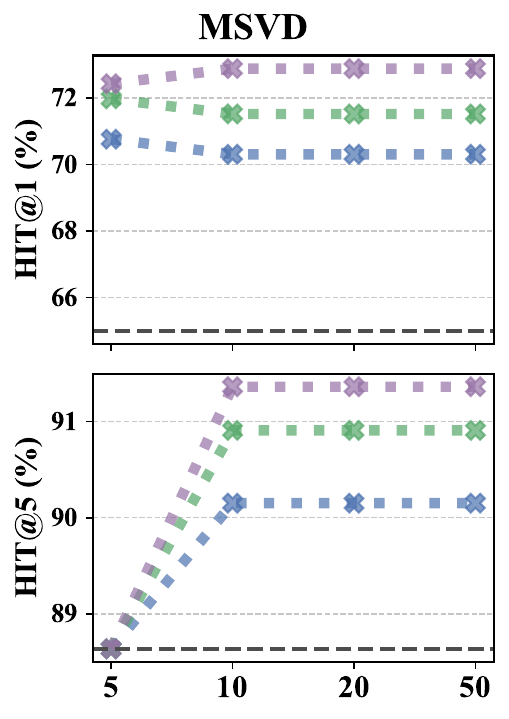}
        % \caption{VQA}
    \end{subfigure}
    \begin{subfigure}[t]{0.15\textwidth}
        \includegraphics[width=\linewidth]{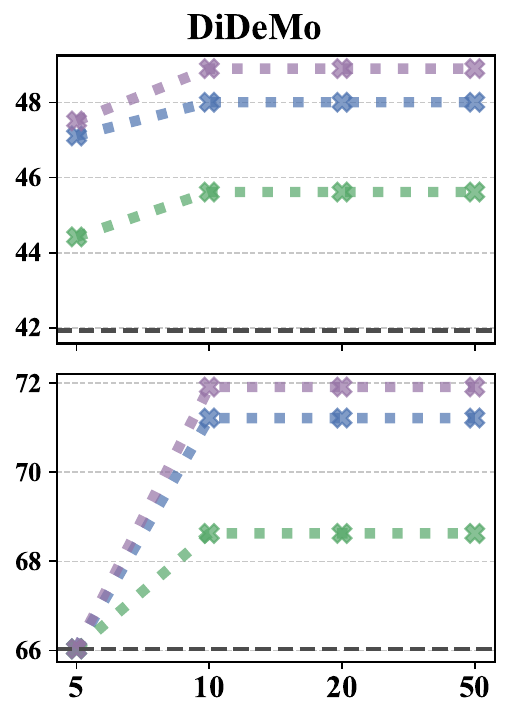}
        % \caption{Grounding}
    \end{subfigure}
    \begin{subfigure}[t]{0.15\textwidth}
        \includegraphics[width=\linewidth]{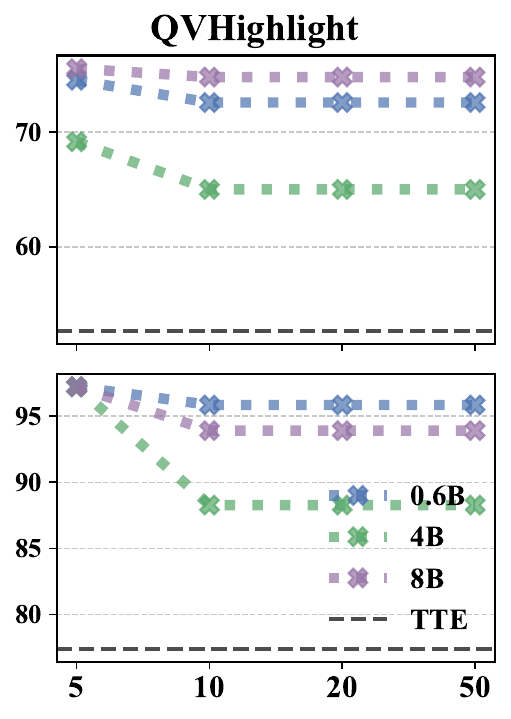}
        % \caption{Grounding}
    \end{subfigure}
    \vspace{-10pt}
    \caption{ Effects of reranker sizes versus top-$k$ candidates in \ecrr on three text-to-video retrieval datasets from MMEB V2~\citep{vlm2vecv2}.}
    \label{fig:rerank_others}

\end{wrapfigure}

\paragraph{Reranking performance on reranker size.} In Figure~\ref{fig:rerank_others} we show the ablations on the impact of reranker size and top-$k$ candidates reranking on the performance of \ecrr on the rest of 3 tested video retrieval datasets. We can observe that (1) increasing reranker size notably improve \ecrr performance, and (2) increasing top-$k$ generally improves \ecrr accuracy. However, we notice that the benefit of scaling up top-$k$ mostly occurs with larger reranker size such as 4B and 8B. For weaker reranker such as Qwen3 0.6B, increasing top-$k$ can inversely impact reranking performance, due to the larger number of distractors on top of limited reranker capacity. Lastly, we notice a noticeable performance degradation on QVHighlight when increasing top-$k$ from $5$ to $10$. This is because the top-$5$ accuracy for QVHighlight is already high. Therefore, increasing $k$ from $5$ to $10$ essentially introduces more distractors, negatively impacting performance.

\begin{table*} 
\centering
\renewcommand{\arraystretch}{1.2}
\resizebox{\textwidth}{!}{

\begin{tabular}{lcccccccc}
\toprule
% & \multicolumn{7}{c}{\textbf{Text-to-Video Retrieval}} & \multirow{2}{*}{\textbf{Overall}} \\
\cmidrule(lr){2-8}
\textbf{Method} &
\textbf{DiDeMo} &
\textbf{MSR-VTT} &
\textbf{MSVD} &
\textbf{VATEX} &
\textbf{YouCook2} &
\textbf{QVHighlight} &
\textbf{Charades-STA} & 
\textbf{Overall} \\ 
\midrule
\multicolumn{9}{c}{\textit{Qwen2.5VL 32B}} \\
\midrule
Baseline                                & 32.6 & 27.5 & 51.4 & 23.1 & 12.3 & 27.5 & 16.4 & 27.3 \\
\rowcolor{gray!10}
\ \ \textit{w/ MLLM R.} (Qwen2VL-7B)            & 28.1 & 24.9 & 48.2 & 18.6 & 11.7 & 12.6 & 10.6 & 22.1\mns{5.2} \\
\ \ \textit{w/ MLLM R.} (Qwen2.5VL 32B)         & 33.8 & 29.1 & 49.7 & 23.5 & 14.6 & 15.4 & 12.3 & 25.5\mns{1.8} \\
\rowcolor{gray!10}
\ \ \textit{w/ MLLM R.} (Qwen2.5VL 72B)         & 35.7 & 30.4 & 52.7 & 24.3 & 17.5 & 15.7 & 12.5 & 27.0\mns{0.3} \\
\ \ \textit{w/} \ecrr                            & 34.6 & 30.2 & 51.3 & 22.4 & 17.2 & 59.3 & 42.5 & 36.8\pls{9.5} \\
\midrule 
\multicolumn{9}{c}{\textit{Qwen2.5VL 72B}} \\
\midrule
Baseline                           & 34.5 & 28.6 & 52.7 & 23.8 & 13.0 & 28.3 & 17.3 & 28.3 \\
\rowcolor{gray!10}
\ \ \textit{w/ MLLM R.} (Qwen2VL-7B)            & 28.7 & 25.7 & 47.4 & 19.4 & 11.4 & 12.7 & 10.6 & 22.3\mns{6.0} \\
\ \ \textit{w/ MLLM R.} (Qwen2.5VL 32B)         & 34.4 & 29.4 & 51.6 & 24.7 & 16.5 & 15.4 & 12.3 & 26.3\mns{2.0} \\
\rowcolor{gray!10}
\ \ \textit{w/ MLLM R.} (Qwen2.5VL 72B)         & 36.2 & 31.7 & 53.2 & 25.2 & 18.3 & 15.8 & 12.6 & 27.6\mns{0.7} \\
\ \ \textit{w/} \ecrr                            & 35.8 & 33.2 & 52.7 & 26.5 & 19.1 & 62.8 & 44.1 & 39.2\pls{10.9} \\
\midrule
\multicolumn{9}{c}{\textit{Gemini2.5Pro}} \\
\midrule
Baseline                           & 35.7 & 33.9 & 61.7 & 31.5 & 18.7 & 40.3 & 21.4 & 34.7 \\
\rowcolor{gray!10}
\ \ \textit{w/ MLLM R.} (Qwen2VL-7B)            & 30.3 & 28.4 & 50.6 & 25.3 & 14.2 & 12.8 & 10.7 & 24.6\mns{10.1} \\
\ \ \textit{w/ MLLM R.} (Qwen2.5VL 32B)         & 35.2 & 34.1 & 56.2 & 27.5 & 17.8 & 15.4 & 12.2 & 28.3\mns{6.4} \\
\rowcolor{gray!10}
\ \ \textit{w/ MLLM R.} (Qwen2.5VL 72B)         & 37.2 & 35.6 & 58.4 & 29.7 & 20.5 & 15.7 & 12.6 & 30.0\mns{4.7} \\
\ \ \textit{w/} \ecrr                            & 47.9 & 47.7 & 72.2 & 39.2 & 30.1 & 66.2 & 49.3 & \textbf{48.6}\pls{\textbf{13.9}} \\
\bottomrule
\end{tabular}

}
\caption{Ablations on the impact of the quality of ECRs on reranking. All experiments are based on Qwen2-VL 2B~\cite{qwen2vl} backbone embedder. Results for \ecrr are obtained with Qwen3-8B Reranker~\citep{qwen3}.}

\label{tab:abl_reranking_ecr_model}
\end{table*}

\begin{table*}[t!]
\centering
\renewcommand{\arraystretch}{1.2}
\resizebox{0.85\linewidth}{!}{

\begin{tabular}{lcccccccc}
\toprule
\textbf{Method} &
\textbf{DiDeMo} &
\textbf{MSR-VTT} &
\textbf{MSVD} &
\textbf{VATEX} &
\textbf{YouCook2} &
\textbf{QVHighlight} &
\textbf{Charades-STA} &
\textbf{Overall} \\
\midrule
Baseline
& 35.7 & 33.9 & 61.7 & 31.5 & 18.7 & 40.3 & 21.4 & 34.7 \\
\rowcolor{gray!10}
\textit{w/ MLLM R.}            
& 37.2 & 35.6 & 58.4 & 29.7 & 20.5 & 15.7 & 12.6 & 30.0\mns{4.7} \\
\midrule
\textit{w/} \ecrr           
& 47.9 & 47.7 & 72.2 & 39.2 & 30.1 & 66.2 & 49.3 & 50.4\pls{15.7} \\
\rowcolor{gray!10}
\ \ \textit{w/} \qar (Qwen2.5-VL 32B)         
& 46.3 & 45.9 & 70.4 & 41.2 & 28.4 & 68.8 & 52.6 & 50.5\pls{15.8} \\
\ \ \textit{w/} \qar (Qwen2.5-VL 72B)       
& 48.7 & 47.5 & 71.8 & 43.5 & 30.6 & 71.3 & 54.8 & 52.6\pls{17.9} \\
\rowcolor{gray!10}
\ \ \textit{w/} \qar (Gemini 2.5Pro)       
& 51.4 & 50.4 & 74.3 & 46.4 & 32.8 & 77.2 & 78.9 & \textbf{58.8}\pls{\textbf{24.1}} \\
\bottomrule
\end{tabular}

}
\caption{Ablations on \qar \textit{w.r.t.} different model sizes for performing \qar, for video retrieval tasks, with top-$10$ reranking candidates. Shorthands: \textbf{MLLM R.}: MLLM-based multimodal reranker. \textbf{L.}: listwise reranking using Gemini2.5Pro; \textbf{P.}: pairwise reranking using Qwen3 8B. Datasets from left to right: DiDeMo, MSR-VTT, MSVD, VATEX, YouCook2, QVHighlight, Charades-STA. All experiments are based on Qwen2-VL 2B~\cite{qwen2vl} backbone embedder.}
\label{tab:abl_qar}
\end{table*}

\section{Qualitative examples}

\begin{figure*}
    \centering
    \includegraphics[width=\linewidth]{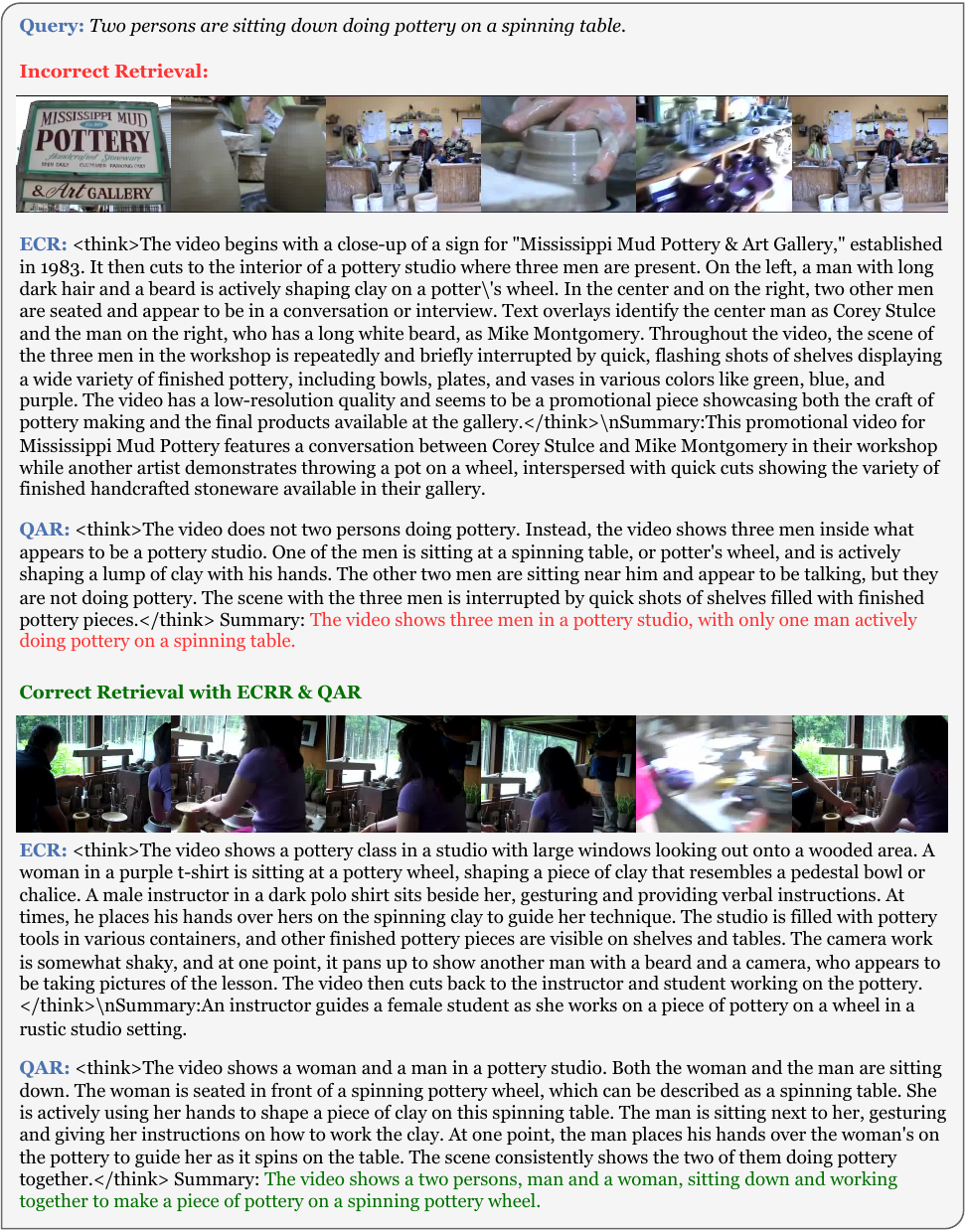}
    % \vspace{-35pt}
    \caption{Qualitative examples for the effect of QAR.}
    \label{fig:qar_examples}
\end{figure*}

In Figure~\ref{fig:qar_examples} we show qualitative examples from text-to-video retrieval task for the effect of \qar. We can observe that \qar brings noticeably more discriminative information relevant to the query. For instance, in the top example, while the original ECR correctly captures the main content of the video, it failed to capture half of the content of the caption (i.e. \textit{``A group of people walking along the street}), while only concentrating on the main character of the video (i.e. the boy). On the other hand, the \qar is able to provide more relevant information \textit{w.r.t.} the query.

\section{Implementation Details}

\paragraph{Training.} We perform training on 64 Nvidia A100 GPUs with Pytorch version \texttt{2.9.0+cu128}. We use FlashAttention2~\citep{flashattn2} for both training and inference. Models are trained for 2.3 epochs with AdamW optimizer, warm-up ratio of 5\%, a global batch size of $8192$, learning rate of $2\times10^{-4}$, and contrastive temperature $\tau=0.02$. We apply LoRA~\citep{lora} fine-tuning with rank 16 and $\alpha=64$, and utilize GradCache~\citep{gradcache} to enable large per-device batches. We use  an interleaved batch size of 64, following VLM2Vec-V2~\citep{vlm2vecv2}. We follow the same system and task prompts as in TTE~\citep{tte}. 

\begin{wrapfigure}{r}{0.5\linewidth}
    \centering
    \includegraphics[width=\linewidth]{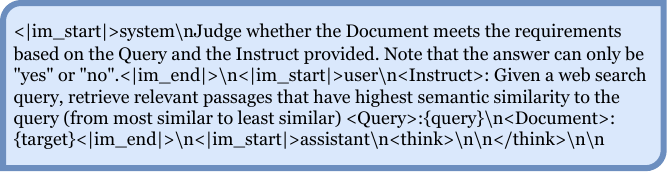}
    \vspace{-15pt}
    \caption{Prompt template for Qwen3 reranker used in our \ecrr.}
    \vspace{-5pt}
    \label{fig:reranker_prompt}
\end{wrapfigure}

\paragraph{Reranking.} For \ecrr, we utilize Qwen3 reranker~\citep{qwen3} as the pairwise reranker. Qwen3 reranker is a text-only reranker. We pass in the query text and each target ECR to the reranker which returns a scalar denoting the matching score. When there are ECRs on the query side (e.g. on moment retrieval, where we apply ECR to both query and target), we pass in the query ECR instead of the original query text to the reranker. The prompt for reranker is shown in Figure~\ref{fig:reranker_prompt}.

We use MLLM-based zero-shot reranking (abbreviated as \textit{MLLM R.}) ~\citep{mm_embed} as the baseline reranking method compared to our proposed \ecrr. This method leverages an off-the-shelf MLLM (e.g. Qwen2.5VL-72B) for reranking, by passing in both query text and candidate video (i.e. for text-to-video), and prompt it to determine how well does the candidate match the query. Following \citealt{mm_embed}, we use the prompt: \texttt{``<video>\textbackslash nQuery: <caption>Does the above video match the caption? yes or no''}. The matching score is calculated by taking the logits of the \texttt{yes} token and normalize it with the logits of \texttt{no} token.

% \newpage

\begin{figure*}
    \centering
    \includegraphics[width=\linewidth]{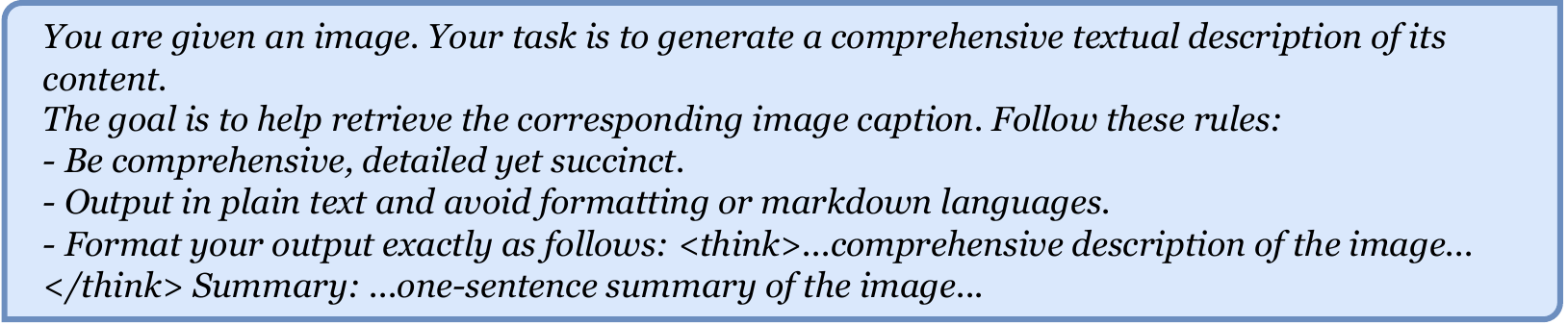}
    \caption{Query-side ECR generation prompt for CrossModal3600~\citep{crossmodal3600} and XTD10~\cite{xtd10}.}
    \label{fig:multilingual_prompt}
\end{figure*}

\begin{figure*}
    \centering
    \includegraphics[width=\linewidth]{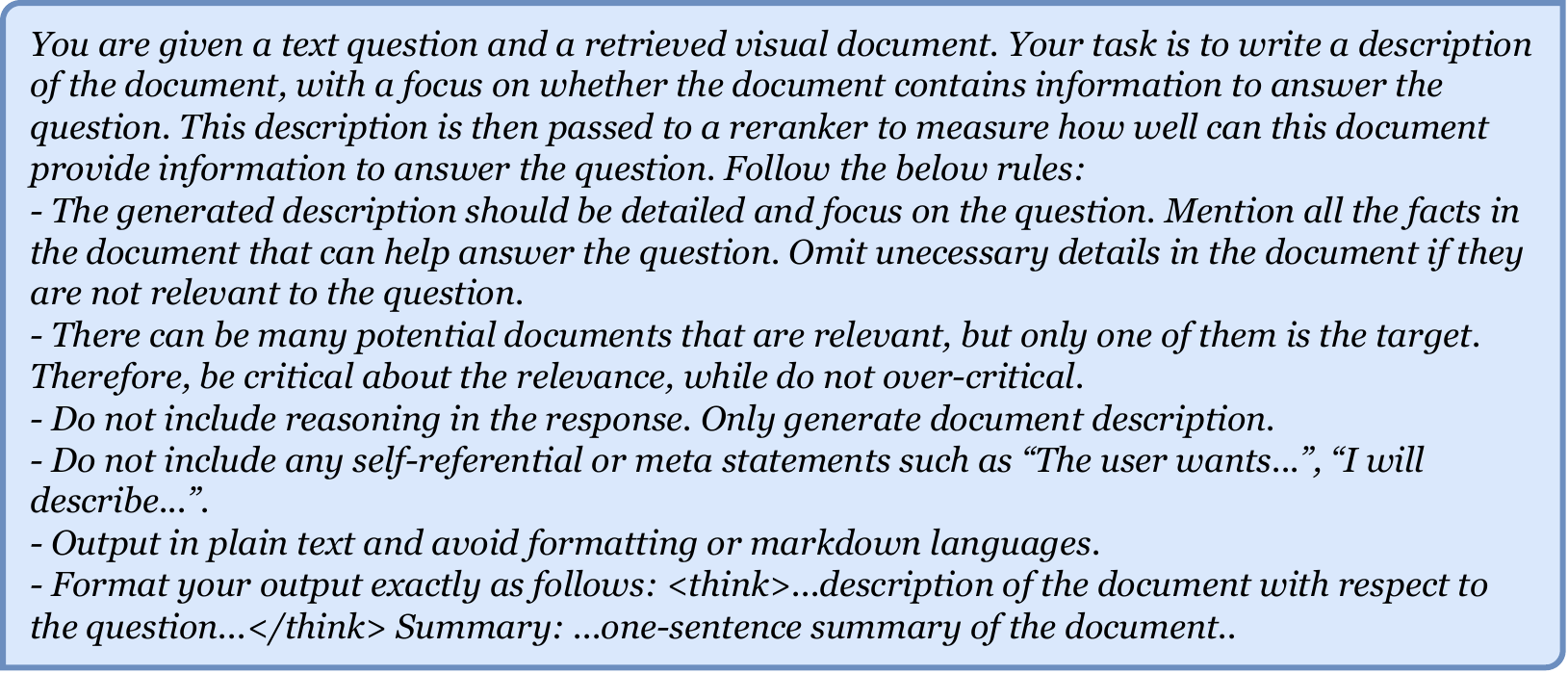}
    \caption{Target-side Query-Aware Reasoning (\qar) generation prompt for video tasks.}
    \label{fig:qar_video_prompt}
\end{figure*}

\begin{figure*}
    \centering
    \includegraphics[width=\linewidth]{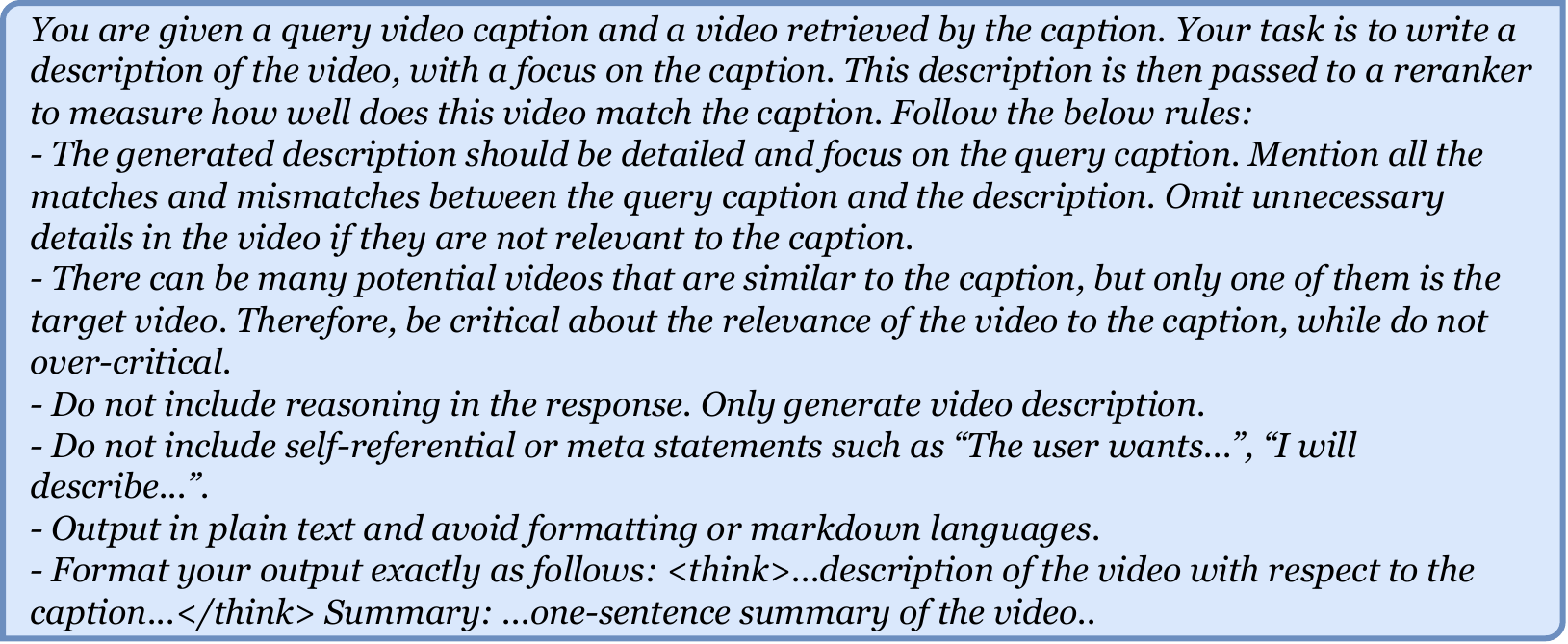}
    \caption{Target-side Query-Aware Reasoning (\qar) generation prompt for visdoc tasks.}
    \label{fig:qar_visdoc_prompt}
\end{figure*}

\begin{figure*}
    \centering
    \includegraphics[width=\linewidth]{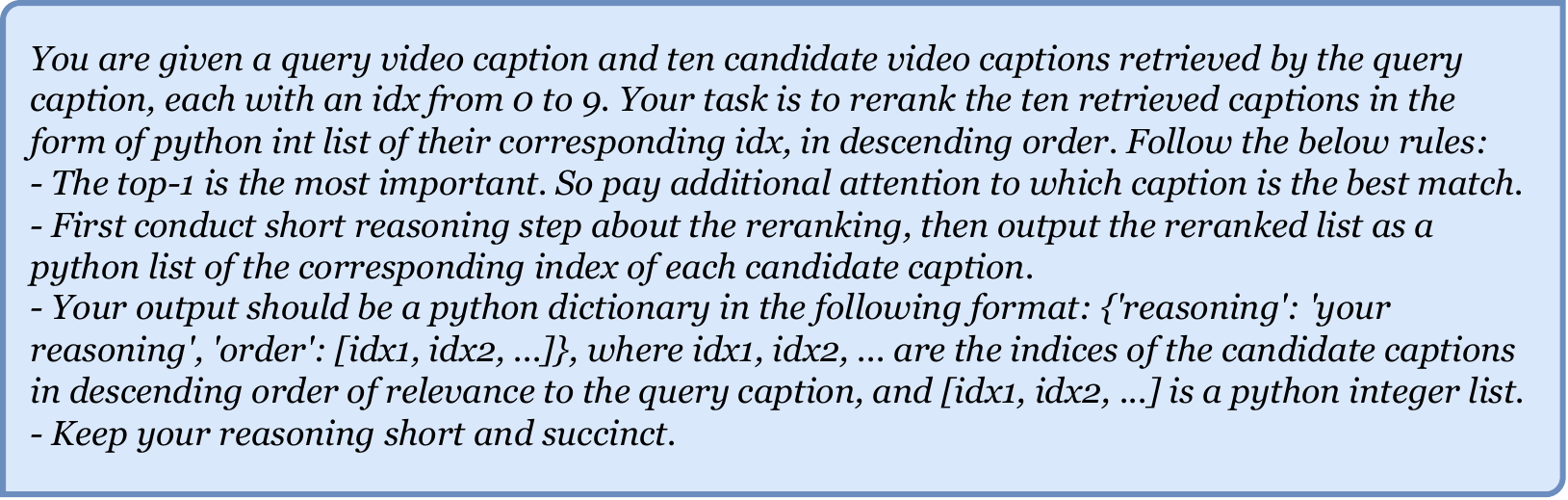}
    \caption{Prompts for list-wise reranking (text-only) using Gemini2.5 Pro~\citep{gemini2_5}}.
    \label{fig:gemini_reranking_prompt}
\end{figure*}

\end{document}